\pdfoutput=1
\RequirePackage{ifpdf}
\ifpdf % We are running pdfTeX in pdf mode
\documentclass[pdftex]{sigma}
\else
\documentclass{sigma}
\fi
\usepackage{enumerate}

\newcommand{\abs}[1]{\left\vert {#1} \right\vert}
\newcommand{\prt}[1]{\left( {#1} \right)}

\newcommand{\scal}[1]{\left< {#1} \right>}
\newcommand{\dv}[2]{\frac{\mathrm d{#1}}{\mathrm d{#2}}}
\newcommand{\pd}[2]{\frac{\partial{#1}}{\partial{#2}}}
\renewcommand{\H}{\mathcal{H}}

\DeclareMathOperator{\Span}{span}
\DeclareMathOperator{\diag}{diag}

\DeclareMathOperator{\tr}{tr}
\DeclareMathOperator{\sgn}{sgn}
\DeclareMathOperator{\Tr}{Tr}
\begin{document}

\allowdisplaybreaks

\renewcommand{\thefootnote}{$\star$}

\renewcommand{\PaperNumber}{010}

\FirstPageHeading

\ShortArticleName{Exploring the Causal Structures of Almost Commutative Geometries}

\ArticleName{Exploring the Causal Structures\\
of Almost Commutative Geometries\footnote{This paper is a~contribution to the Special Issue on
Noncommutative Geometry and Quantum Groups in honor of Marc A.~Rief\/fel.
The full collection is available at
\href{http://www.emis.de/journals/SIGMA/Rieffel.html}{http://www.emis.de/journals/SIGMA/Rief\/fel.html}}}   

\Author{Nicolas FRANCO~$^\dag$ and Micha{\l} ECKSTEIN~$^{\ddag\dag}$}

\AuthorNameForHeading{N.~Franco and M.~Eckstein}

\Address{$^\dag$~Copernicus Center for Interdisciplinary Studies, ul.~S{\l}awkowska 17, 31-016 Krak\'ow, Poland}
\EmailD{\href{mailto:nicolas.franco@math.unamur.be}{nicolas.franco@math.unamur.be},
\href{mailto:nicolas.franco@im.uj.edu.pl}{nicolas.franco@im.uj.edu.pl}}

\Address{$^\ddag$~Faculty of Mathematics and Computer Science, Jagellonian University,\\
\hphantom{$^\ddag$}~ul.~{\L}ojasiewicza 6, 30-348 Krak\'ow, Poland}
\EmailD{\href{mailto:michal.eckstein@uj.edu.pl}{michal.eckstein@uj.edu.pl}}

\ArticleDates{Received October 31, 2013, in f\/inal form January 20, 2014; Published online January 28, 2014}

\Abstract{We investigate the causal relations in the space of states of almost commutative Lorentzian
geometries.
We fully describe the causal structure of a~simple model based on the algebra
$\mathcal{S}(\mathbb{R}^{1,1}) \otimes M_2(\mathbb{C})$, which has a~non-trivial space of internal degrees
of freedom.
It turns out that the causality condition imposes restrictions on the motion in the internal space.
Moreover, we show that the requirement of causality favours a~unitary evolution in the internal space.}

\Keywords{noncommutative geometry; causal structures; Lorentzian spectral triples}

\Classification{58B34; 53C50; 54F05}

\renewcommand{\thefootnote}{\arabic{footnote}} \setcounter{footnote}{0}

\section{Introduction}

Noncommutative geometry \`a la Connes~\cite{C94} of\/fers new, vast horizons for both pure mathe\-ma\-tics, as
well as physical models.
In its original formulation it provides a~signif\/icant generalisation of Riemannian dif\/ferential
geometry.
However, for physical applications one should prefer Lorentzian rather than Riemannian structures.
This is especially relevant for applications related to gravitation.
To this end the theory of Lorentzian spectral triples, a~generalisation of the notion of spectral triples
based on Krein space approach, has been developed by several
authors~\cite{F4,F5,F6,Pas,Stro,Rennie12,Suij,Verch11}.
We can also mention several applications of noncommutative geometry in physics and cosmology in
a~Lorentzian setting, that use other techniques (for instance Wick
rotation)~\cite{Barret,BCSV,BDRS2,BDRS,MPT,Mor,Sakellariadou}.

One of the most important elements of Lorentzian geometry, that is absent in Riemannian theory, is the
causal structure.
For a~mathematician this is a~partial order relation def\/ined between the points of a~Lorentzian manifold.
On the other hand, it is of crucial important for physicists since it determines which of the space-time
events can be linked by a~physical (not faster than light) signal.

The generalisation of the causal structure to noncommutative spaces is not at all straightforward because
of the lack of the notion of points (or events).
One of the possible counterparts of points in noncommutative geometry are pure states on algebras.
By the famous Gelfand--Naimark theorem, pure states on a~commutative $C^*$-algebra $\mathcal{A}$ are in one
to one correspondence with points of some locally compact Hausdorf\/f space $\mathrm{Spec}(\mathcal{A})$.
In~\cite{F6} we proposed a~def\/inition of a~causal structure for the space of states in the general
framework of Lorentzian spectral triples.
We have succeeded in showing that the usual causal relations are recovered whenever the spectral triple is
commutative and constructed from a~globally hyperbolic Lorentzian manifold.

The purpose of this paper is to investigate the properties of such generalised causal structures for
a~class of noncommutative spaces described by almost commutative geometries.
The latter consists in considering spectral triples based on noncommutative algebras of the form
\mbox{${C}^{\infty}(\mathcal{M}) \otimes \mathcal{A}_F$},
with $\mathcal{A}_F$ being a~f\/inite direct sum
of matrix algebras.
Almost commutative geometries are of crucial importance in physical applications of noncommutative geometry
as they provide a~f\/irm framework for models of fundamental interactions~\cite{CCM06,C94,MC08,Dungen}.

The plan of the paper presents itself as follows.
Section~\ref{General} contains some general results on almost commutative Lorentzian geometries and
their causal structure.
Later, in Section~\ref{Example}, we present in details the causal structure of a~particular spectral
triple based on the algebra $\mathcal{S}({\mathbb R}^{1,1}) \otimes M_2({\mathbb C})$.
We unravel the causal relations between the pure states (Section~\ref{secps}) as well as for
a~preferred subclass of mixed states (Section~\ref{secms}).
We end with an outlook for further exploration of causal structures of almost commutative geometries along
with their physical interpretation.

\section{Lorentzian almost commutative geometries\\
and causal structures}
\label{General}

The basic objects of Riemannian noncommutative geometry are spectral triples~\cite{C94,MC08}.
Their generalisation to a~pseudo-Riemannian setting is based on a~working set of axioms, which slightly
varies among the dif\/ferent approaches~\cite{F4,F5,Pas,Rennie12,Verch11}.
We base our work essentially on the formulation adopted in~\cite{F4,F5,F6}.
However, our results on causal structure can be accommodated also in the other approaches as already argued
in~\cite{F6}.
\begin{definition}
\label{deflost}
A~Lorentzian spectral triple is given by the data $(\mathcal{A},\widetilde{\mathcal{A}},\H,D,\mathcal{J})$
with:
\begin{itemize}
\itemsep=0pt \item A~Hilbert space $\H$.
\item A~non-unital pre-$C^*$-algebra $\mathcal{A}$, with a~faithful representation as bounded operators on
$\H$.
\item A~preferred unitisation $\widetilde{\mathcal{A}}$ of $\mathcal{A}$, which is also a~pre-$C^*$-algebra
with a~faithful representation as bounded operators on $\H$ and such that $\mathcal{A}$ is an ideal of
$\widetilde{\mathcal{A}}$.
\item An unbounded operator $D$, densely def\/ined on $\H$, such that:
\begin{itemize}\itemsep=0pt
\item $\forall\, a\in\widetilde{\mathcal{A}}$,   $[D,a]$ extends to a~bounded operator on $\H$, \item $\forall\,
a\in\mathcal{A}$,   $a(1 + \scal{D}^2)^{-\frac 12}$ is compact, with $\scal{D}^2 = \frac 12 (D D^* + D^*
D)$.
\end{itemize}
\item A~bounded operator $\mathcal{J}$ on $\H$ such that:
\begin{itemize}\itemsep=0pt
 \item $\mathcal{J}^2=1$, \item $\mathcal{J}^*=\mathcal{J}$, \item $[\mathcal{J},a]=0$,
$\forall\, a\in\widetilde{\mathcal{A}}$, \item $D^*=-\mathcal{J} D \mathcal{J}$, \item
$\mathcal{J}  = -[D,\mathcal{T}]$ for some unbounded self-adjoint operator $\mathcal{T}$ with domain
\mbox{$\text{Dom}(\mathcal{T})  \subset  \H$}, such that  $\prt{1+ \mathcal{T}^2}^{-\frac{1}{2}}\in
\widetilde{\mathcal{A}}$.
\end{itemize}
\end{itemize}
\end{definition}

The role of the operator $\mathcal{J}$, called fundamental symmetry, is to turn the Hilbert space $\H$ into
a~Krein space on which the operator $iD$ is self-adjoint~\cite{Bog,Stro}.
In the commutative case the condition $\mathcal{J}=-[D,\mathcal{T}]$ guarantees the correct signature, but
restricts the admissible Lorentzian manifolds to the ones equipped with a~global time function~\cite{F5}.
It is suf\/f\/icient for the purposes of~this paper since the considered almost commutative geometries are
based on globally hyperbolic space-times.
For a~more general def\/inition of $\mathcal{J}$ suitable in the pseudo-Riemannian context see~\cite{Pas}
or~\cite{Rennie12}.
\begin{definition}
A~Lorentzian spectral triple is even if there exists a~$\mathbb Z_2$-grading $\gamma$ such that
$\gamma^*=\gamma$, $\gamma^2=1$, $[\gamma,a] = 0$ $\forall\, a\in\widetilde{\mathcal{A}}$, $\gamma
\mathcal{J} =- \mathcal{J} \gamma$ and $\gamma D =- D \gamma $.
\end{definition}

Let us now consider a~locally compact complete globally hyperbolic Lorentzian manifold $\mathcal{M}$ of
dimension $n$ with a~spin structure $S$.
By a~complete Lorentzian manifold we understand the following: there exists a~spacelike ref\/lection~--
i.e.~an automorphism $r$ of the tangent bundle respecting $r^2=1$, $g(r\cdot,r\cdot) = g(\cdot,\cdot)$~--
such that $\mathcal{M}$ equipped with a~Riemannian metric $g^r(\cdot,\cdot) = g(\cdot,r\cdot)$ is complete
in the usual Lebesgue sense.
Given a~globally hyperbolic Lorentzian manifold~$\mathcal{M}$ one can construct a~commutative Lorentzian
spectral triple in the following way~\cite{F6}:
\begin{itemize}\itemsep=0pt
\item $\H = L^2(\mathcal{M},S)$ is the Hilbert space of square integrable sections of the
spinor bundle over~$\mathcal{M}$.
\item $D = -i(\hat c \circ \nabla^S) = -i e^\mu_a\gamma^{a} \nabla^S_\mu$ is the Dirac operator associated
with the spin connection~$\nabla^S$ (the Einstein summation convention is in use and $e^\mu_a$
stand for vielbeins).
\item $\mathcal{A} \subset C^\infty_0(\mathcal{M})$ and $\widetilde{\mathcal{A}} \subset
C^\infty_b(\mathcal{M})$ with pointwise multiplication are some appropriate sub-algebras of the algebra of
smooth functions vanishing at inf\/inity and the algebra of smooth bounded functions respectively.
{}$\widetilde{\mathcal{A}}$ must be such that $\forall\, a\in\widetilde{\mathcal{A}}$, $[D,a]$ extends to
a~bounded operator on $\H$.
The representation is given by standard multiplication operators on $\H$: $(\pi(a) \psi)(x) = a(x) \psi(x)$
for all $x \in \mathcal{M}$.
\item $\mathcal{J}=i\gamma^0$, where $\gamma^0$ is the f\/irst f\/lat gamma matrix\footnote{Conventions
used in the paper are $(-,+,+,+,\cdots)$ for the signature of the metric and
$\{\gamma^a,\gamma^b\}=2\eta^{ab}$ for the f\/lat gamma matrices, with $\gamma^0$ anti-Hermitian and
$\gamma^a$ Hermitian for $a>0$.}.
\end{itemize}
If $n$ is even, the $\mathbb Z_2$-grading is given by the chirality element: $\gamma = (-i)^{\frac{n}{2} +
1} \gamma^0 \cdots \gamma^{n-1}.$

An even Lorentzian spectral triple can be combined with a~Riemannian spectral triple in order to obtain
a~new Lorentzian spectral triple.
Only one of the two spectral triples should be Lorentzian, otherwise the obtained spectral triple would be
pseudo-Riemannian with a~more general signature (as def\/ined in~\cite{Stro}).
When the Lorentzian spectral triple is commutative and the Riemanian one is noncommutative and f\/inite, we
obtain an almost commutative geometry with Lorentzian signature.
\begin{theorem}
\label{ProductThm}
Let us consider an even Lorentzian spectral triple
$(\mathcal{A}_\mathcal{M},\widetilde{\mathcal{A}}_\mathcal{M},\H_\mathcal{M},D_\mathcal{M},\mathcal{J}_\mathcal{M})$
with $\mathbb Z_2$-grading $\gamma_\mathcal{M}$ and a~$($compact$)$ Riemannian spectral triple $(\mathcal{A}_F,\H_F,D_F)$,
then the product
\begin{gather*}
\mathcal{A}=\mathcal{A}_\mathcal{M}\otimes\mathcal{A}_F,
\\
\widetilde{\mathcal{A}}=\widetilde{\mathcal{A}}_\mathcal{M}\otimes\mathcal{A}_F,
\\
H=H_\mathcal{M}\otimes H_F,
\\
D=D_\mathcal{M}\otimes1+\gamma_\mathcal{M}\otimes D_F,
\\
\mathcal{J}=\mathcal{J}_\mathcal{M}\otimes1
\end{gather*}
is a~Lorentzian spectral triple.
\end{theorem}
\begin{proof}
Since the above def\/inition is completely analogous to the standard product between two Riemannian
spectral triples~\cite{C94,MC08}, we only need to show that the conditions related to the fundamental
symmetry $\mathcal{J}$ are fulf\/illed.

The conditions $\mathcal{J}^2=1$, $\mathcal{J}^*=\mathcal{J}$ and $[\mathcal{J},a]=0$ $\forall\,
a\in\widetilde{\mathcal{A}}$ are easily derived from the properties $\mathcal{J}_\mathcal{M}^2=1$,
$\mathcal{J}_\mathcal{M}^*=\mathcal{J}_\mathcal{M}$ and $[\mathcal{J}_\mathcal{M},a]=0$ $\forall\,
a\in\widetilde{\mathcal{A}}_\mathcal{M}$.

The Hermicity condition of the Dirac operator is met:
\begin{gather*}
\mathcal{J}D\mathcal{J}=\mathcal{J}_\mathcal{M}D_\mathcal{M}\mathcal{J}_\mathcal{M}\otimes1+\mathcal{J}
_\mathcal{M}\gamma_\mathcal{M}\mathcal{J}_\mathcal{M}\otimes D_F=-D^*_\mathcal{M}
\otimes1^*-\gamma^*_\mathcal{M}\otimes D^*_F=-D^*,
\end{gather*}
since $ \mathcal{J}_\mathcal{M}\gamma_\mathcal{M} \mathcal{J}_\mathcal{M}=-\mathcal{J}^2_\mathcal{M}
\gamma_\mathcal{M} = -\gamma_\mathcal{M} = -\gamma^*_\mathcal{M}$ and $D_F$ is self-adjoint.

We have $\mathcal{J}_\mathcal{M} = -[D,\mathcal{T}_M]$ for some unbounded self-adjoint operator
$\mathcal{T}_\mathcal{M}$ (which commutes with $\gamma_\mathcal{M}$), so by setting $\mathcal{T} =
\mathcal{T}_\mathcal{M} \otimes 1$ we f\/ind:
\begin{gather*}
-[D,\mathcal{T}]=-D_\mathcal{M}\mathcal{T}_\mathcal{M}\otimes1-\gamma_\mathcal{M}\mathcal{T}
_\mathcal{M}\otimes D_F+\mathcal{T}_\mathcal{M}D_\mathcal{M}\otimes1+\mathcal{T}_\mathcal{M}
\gamma_\mathcal{M}\otimes D_F
\\
\phantom{-[D,\mathcal{T}]}{}
=-[D_\mathcal{M},T_\mathcal{M}]\otimes1=\mathcal{J}_\mathcal{M}\otimes1=\mathcal{J}.
\end{gather*}
Hence, the resulting spectral triple is indeed of the Lorentzian signature.
\end{proof}

Let us remark that, if the Riemannian spectral triple is even with $\mathbb Z_2$-grading $\gamma_F$, then
the product can also be def\/ined with an alternative operator $D = D_\mathcal{M} \otimes \gamma_F + 1
\otimes D_F$.
In this case the new fundamental symmetry must be set to $\mathcal{J} = \mathcal{J}_\mathcal{M} \otimes
\gamma_F$.
If both spectral triples are even, then the product is even with the $\mathbb Z_2$-grading $\gamma =
\gamma_\mathcal{M} \otimes \gamma_F$.

To make the paper self-contained we shall recollect the basic def\/initions and properties concerning the
causal structure for Lorentzian spectral triples.
As mentioned in the Introduction, the causal relations are def\/ined between the states on the algebra~$\widetilde{\mathcal{A}}$, which can be viewed as a~noncommutative generalisation of events (points of
space-time).
Let us recall that a~(mixed) state on a~$C^*$-algebra is a~positive linear functional (automatically
continuous) of norm one~\cite{Kadison}.
For our purposes we consider the states on the algebra $\overline{\widetilde{\mathcal{A}}}$~-- the
$C^*$-completion of $\widetilde{\mathcal{A}}$~-- restricted to $\widetilde{\mathcal{A}}$.
We shall denote the set of such states as $S(\widetilde{\mathcal{A}})$.
It is a~closed convex set for the weak-$^*$ topology and its extremal points form a~set of pure states
denoted by~$P(\widetilde{\mathcal{A}})$.

The causal relation between the states is def\/ined in the following way~\cite{F6}:
\begin{definition}

\label{defcausal}
Let us consider the cone $\mathcal{C}$ of all Hermitian elements $a \in \widetilde{\mathcal{A}}$ respecting
\begin{gather}
\label{causality1}
\forall\,\phi\in\H
\qquad
\scal{\phi,\mathcal{J}[D,a]\phi}\leq0,
\end{gather}
where $\scal{\cdot,\cdot}$ is the inner product on $\H$.
If the following condition is fulf\/illed:
\begin{gather}
\label{causality2}
\overline{\Span_{{\mathbb C}}(\mathcal{C})}=\overline{\widetilde{\mathcal{A}}},
\end{gather}
then $\mathcal{C}$ is called a~\emph{causal cone} and def\/ines a~partial order relation on
$S(\widetilde{\mathcal{A}})$ by
\begin{gather*}
\forall\, \omega,\eta\in S(\widetilde{\mathcal{A}})
\quad
\omega\preceq\eta
\qquad
\text{if\/f}
\qquad
\forall\, a\in\mathcal{C}
\quad\omega(a)\leq\eta(a).
\end{gather*}
\end{definition}

Since the def\/inition of a~Lorentzian spectral triple allows for a~preferred choice of the unitisation
$\widetilde{\mathcal{A}}$, one may actually use the condition~\eqref{causality2} to determine it,
as~\eqref{causality2} may, in general, be false for an arbitrary unitisation.
We shall use the following procedure in order to determine the preferred unitisation:
\begin{enumerate}
\itemsep=0pt \item Chose $\widetilde{\mathcal{A}}_0$ as the biggest possible unitisation fulf\/illing the
axioms of Def\/inition~\ref{deflost}.
\item Def\/ine $\mathcal{C}$ to be the set of all of the elements in $\widetilde{\mathcal{A}}_0$
fulf\/illing the condition~\eqref{causality1}.
\item Set $\widetilde{\mathcal{A}} = \Span_{{\mathbb C}}(\mathcal{C})$ and check that it is a~valid
unitisation for $\mathcal{A}$.
\end{enumerate}

This procedure assures that $\mathcal{C}$ is a~causal cone which is automatically the maximal one.
The latter is important, since in the commutative case the maximal causal cone is needed to recover the
classical causal structure on space-time~\cite{F6}.
In the case of a~globally hyperbolic Lorentzian manifold, such a~suitable unitisation always
exists~\cite{Bes,F6}.
We shall see that for almost commutative geometries the described procedure is also sound.

Motivated by the following theorem, we call the partial order relation $\preceq$ on
$S(\widetilde{\mathcal{A}})$ a~\emph{causal relation} between the states.
\begin{theorem}[\protect{\cite{F6}}]
\label{thmreconstruction}
Let $(\mathcal{A},\widetilde{\mathcal{A}},\H,D,\mathcal{J})$ be a~commutative Lorentzian spectral
triple constructed from a~globally hyperbolic Lorentzian manifold $\mathcal{M}$, and let us define the
following subset of pure states:
\begin{gather*}
\mathcal{M}(\widetilde{\mathcal{A}})
=\big\{\omega\in P(\widetilde{\mathcal{A}}):\mathcal{A}\not\subset\ker\omega\big\}\subset S(\widetilde{\mathcal{A}}).
\end{gather*}
Then the causal relation $\preceq$ on $S(\widetilde{\mathcal{A}})$ restricted to
$\mathcal{M}(\widetilde{\mathcal{A}}) \cong \mathcal{M}$ corresponds to the usual causal relation on
$\mathcal{M}$.
\end{theorem}

The proof is based on the one-to-one correspondence between the elements of $\mathcal{C}$ and the causal
functions, which are the real-valued functions on $\mathcal{M}$ non-decreasing along future directed causal
curves.
This theorem shows the necessity of adding a~unitisation in the def\/inition of a~Lorentzian spectral
triple, since the subset of causal functions within the initial non-unital algebra only contains the null
function which is not suf\/f\/icient to characterise causality.
The role of the restriction of the space of states to $\mathcal{M}(\widetilde{\mathcal{A}})$ is to avoid the
states localised at inf\/inity.
In fact
we have
$\mathcal{M}(\widetilde{\mathcal{A}}) \cong P(\mathcal{A})$ which, by the Gelfand--Naimark
theorem, is isomorphic to the locally compact manifold $\mathcal{M}$ itself.
The complete proof of this theorem, as well as more details about the def\/inition of causality in the
space of states, can be found in $\cite{F6}$.

Let us now take a~look at the space of states of almost commutative geometries.
In full analogy to the Riemannian case (see for instance~\cite{Dungen}) we def\/ine an \emph{almost
commutative Lorentzian spectral triple} as a~product def\/ined in Theorem~\ref{ProductThm} with
$(\mathcal{A}_\mathcal{M},\widetilde{\mathcal{A}}_\mathcal{M},\H_\mathcal{M},D_\mathcal{M},\mathcal{J}_\mathcal{M})$
constructed from a~globally hyperbolic manifold and $(\mathcal{A}_F,\H_F,D_F)$
based on a~f\/inite direct sum of matrix algebras.
\begin{proposition}
\label{PureStates}
Let $(\mathcal{A},\widetilde{\mathcal{A}},\H,D,\mathcal{J})$ be an almost commutative Lorentzian spectral
triple and let $P(\mathcal{A})$ denote the space of pure states $($extremal points of $S(\mathcal{A}))$ on
$\mathcal{A}$.
Then
\begin{gather*}
P(\widetilde{\mathcal{A}})\cong P(\widetilde{\mathcal{A}}_\mathcal{M})\times P(\mathcal{A}_F),
\qquad
P(\mathcal{A})\cong\mathcal{M}\times P(\mathcal{A}_F).
\end{gather*}
\end{proposition}
\begin{proof}
This facts follows from~\cite[Theorem 11.3.7]{Kadison} (see also~\cite{AndreaMartinetti}) and the locally
compact version of the Gelfand--Naimark theorem.
\end{proof}

This motivates us to def\/ine the following space of \emph{physical states} of almost commutative
Lorentzian geometries
\begin{gather}
\label{MA}
\mathcal{M}(\widetilde{\mathcal{A}})
=\big\{\chi\otimes\xi\,\big\vert\,\chi\in P(\widetilde{\mathcal{A}}_\mathcal{M}),
\;\mathcal{A}_\mathcal{M}\not\subset\ker\chi,\;\xi\in P(\mathcal{A}_F)\big\}\cong P(\mathcal{A}).
\end{gather}

Let us stress that the fact that all of the pure states on $\widetilde{\mathcal{A}}$ are separable
(i.e.~have the form of simple tensor product) is a~consequence of $\mathcal{A}_\mathcal{M}$ being
commutative.
Hence, it is a~genuine property of all almost commutative geometries.
On the other hand, the space of all (mixed) states~$S(\widetilde{\mathcal{A}})$ is far more complicated to
determine.
We shall not attempt to address this problem here, some results on the causal structure of a~subspace of
mixed states are presented in Section~\ref{secms}.

Proposition~\ref{PureStates} has a~useful consequence, which simplif\/ies the examination of the causal
structure and remains valid for all almost commutative Lorentzian geometries.
\begin{proposition}
\label{Unitary}
Let $(\mathcal{A},\widetilde{\mathcal{A}},\H,D,\mathcal{J})$ be an almost commutative Lorentzian spectral
triple and let $\widetilde{D} = D_{\mathcal{M}} \otimes 1 + \gamma_{\mathcal{M}}\otimes \widetilde{D}_F$, with $\widetilde{D}_F = U D_F U^*$, be a~unitary transformation of the finite-part Dirac operator.
If we denote by~$\preceq$ and~$\widetilde\preceq$ the partial order relations associated with the basic and
transformed Dirac operators respectively, then for any $\omega, \eta \in \mathcal{M}(\widetilde{\mathcal{A}})$
\begin{gather*}
\omega\preceq\eta
\quad
\Longleftrightarrow
\quad
(1\otimes U)\omega \; \widetilde\preceq \; (1\otimes U)\eta.
\end{gather*}
\end{proposition}
\begin{proof}
On the strength of Proposition~\ref{PureStates} any state $\omega \in \mathcal{M}(\widetilde{\mathcal{A}})$
can be uniquely written as $\omega = \omega_{p,\xi} = \omega_p \otimes \omega_{\xi}$ with $\omega_p \in
P(\mathcal{A}_\mathcal{M}) \cong \mathcal{M}$ and $\omega_{\xi} \in P(\mathcal{A}_F)$.
Moreover, since $\mathcal{A}_F$ is a~f\/inite direct sum of matrix algebras, every state in
$P(\mathcal{A}_F)$ is a~vector state~\cite{IochKM}, i.e.
\begin{gather*}
\forall\,\omega_{\xi}\in P(\mathcal{A}_F)
\quad
\exists\,\xi\in\H_F \scal{\xi,\xi}=1:
\quad
\forall\, a\in\mathcal{A}
_F,
\quad
\omega_{\xi}(a)=\scal{\xi,a\xi}.
\end{gather*}
As a~digression let us note that if $\mathcal{A}_F$ is not a~single matrix algebra but a~direct product,
then not every normalised vector in $\H_F$ def\/ines a~pure state on $\mathcal{A}_F$.
Now, we have
\begin{gather*}
\phantom{\Longleftrightarrow\quad} \ \forall\, a\in\widetilde{\mathcal{C}},
\quad
(\omega_{p,U\xi}-\omega_{q,U\varphi})(a)\leq0
\\
\Longleftrightarrow \quad \forall\, a\in\widetilde{\mathcal{C}},
\quad
(\omega_{p,\xi}-\omega_{q,\varphi})\big((1\otimes U^*)a(1\otimes U)\big)\leq0,
\end{gather*}
where $\widetilde{\mathcal{C}}$ denotes the causal cone associated with the Dirac operator $\widetilde D$.
The causality condition reads
\begin{gather*}
\phantom{\Longleftrightarrow\quad} \ \forall\, \phi\in\H,
\quad
\big\langle\phi,\mathcal{J}[\widetilde D,a]\phi\big\rangle\leq0
\\
\Longleftrightarrow \quad \forall\,\phi\in\H,
\quad
\scal{\phi,(\mathcal{J}\otimes1)[D_\mathcal{M}\otimes1+\gamma_\mathcal{M}\otimes U D_F U^*,a]\phi}\leq0
\\
\Longleftrightarrow \quad \forall\,\phi\in\H,
\quad
\scal{\phi,(\mathcal{J}\otimes1)[(1\otimes U)D\big(1\otimes U^*\big),a]\phi}\leq0
\\
\Longleftrightarrow \quad \forall\,\phi\in\H,
\quad
\scal{\phi,(\mathcal{J}\otimes1)(1\otimes U)[D,\big(1\otimes U^*\big)a(1\otimes U)](1\otimes U^*)\phi}\leq0
\\
\Longleftrightarrow \quad \forall\,\widetilde\phi=\big(1\otimes U^*\big)\phi\in\H,
\quad
\big\langle\widetilde\phi,(\mathcal{J}\otimes1)[D,\big(1\otimes U^*\big)a(1\otimes U)]\widetilde\phi\big\rangle\leq0
\\
\Longleftrightarrow \quad \big(1\otimes U^*\big)a(1\otimes U)\in\mathcal{C}.
\end{gather*}
Thus, we have
\begin{gather*}
\phantom{\Longleftrightarrow\quad} \  (1\otimes U)\omega_{p,\xi} \; \widetilde\preceq \;  (1\otimes U)\omega_{q,\varphi}
\\
\Longleftrightarrow \quad \forall\, a\in\widetilde{\mathcal{C}},
\quad
(\omega_{p,\xi}-\omega_{q,\varphi})\big((1\otimes U^*)a(1\otimes U)\big)\leq0
\\
\Longleftrightarrow \quad \forall\, \big(1\otimes U^*\big)a(1\otimes U)\in\mathcal{C},
\quad
(\omega_{p,\xi}-\omega_{q,\varphi})\big((1\otimes U^*)a(1\otimes U)\big)\leq0
\\
\Longleftrightarrow \quad \omega_{p,\xi} \preceq \omega_{q,\varphi}.
\tag*{\qed}
\end{gather*}
\renewcommand{\qed}{}
\end{proof}

We conclude this section by investigating the impact of conformal transformations in almost
commutative geometries.
It is a~well known fact that a~conformal rescaling of the metric on a~Lorentzian manifold does not change
its causal structure.
It turns out that for this result to hold true in the almost commutative case one needs to transform the
Dirac operator $D_F$ accordingly.
\begin{proposition}
\label{conformal}
Let $\Omega$ be a~positive smooth bounded function on $\mathcal{M}$ and let
$(\mathcal{A},\widetilde{\mathcal{A}},\H,D,\mathcal{J})$ be an almost commutative Lorentzian spectral triple.
The causal structures of $(\mathcal{A},\widetilde{\mathcal{A}},\H,D,\mathcal{J})$ and
$(\mathcal{A},\widetilde{\mathcal{A}},\H,\widetilde D,\mathcal{J})$, with $\widetilde D = \Omega D \Omega$,
are isomorphic, i.e.
\begin{gather*}
\forall \, \omega, \eta \in S(\widetilde{\mathcal{A}}), \qquad \omega \preceq \eta \quad \Longleftrightarrow \quad \omega \; \widetilde\preceq\; \eta,
\end{gather*}
where $\preceq$ and $\widetilde\preceq$ denote the causal relations on $S(\widetilde{\mathcal{A}})$
associated with $D$ and $\widetilde D$ respectively.
\end{proposition}
\begin{proof}
Let us f\/irst recall that a~conformal rescaling of the metric tensor on $\mathcal{M}$, $\widetilde g =
\Omega^4 g$, results in the change of the Dirac operator of the form $\widetilde D_\mathcal{M} = \Omega
D_\mathcal{M} \Omega$~\cite{Nakahara}.
The condition for an element $a \in \widetilde{\mathcal{A}}$ to be
in the causal cone $\widetilde{\mathcal{C}}$ associated with $\widetilde{D}$ reads
\begin{gather*}
\phantom{\Longleftrightarrow\quad} \ \forall\, \phi\in\H,
\quad
\big\langle\phi,\mathcal{J}[\widetilde D,a]\phi\big\rangle\leq0
\\
\Longleftrightarrow \quad \forall\,\phi\in\H,
\quad
\scal{\phi,\mathcal{J}[\Omega(D_\mathcal{M}\otimes1+\gamma_\mathcal{M}\otimes D_F)\Omega,a]\phi}\leq0
\\
\Longleftrightarrow \quad \forall\,\phi\in\H,
\quad
\scal{\Omega\phi,\mathcal{J}[D_\mathcal{M}\otimes1+\gamma_\mathcal{M}\otimes D_F,a]\Omega\phi}\leq0
\\
\Longleftrightarrow \quad \forall\,\widetilde\phi=\Omega\phi\in\H,
\quad
\big\langle\widetilde\phi,\mathcal{J}[D_\mathcal{M}\otimes1+\gamma_\mathcal{M}\otimes D_F,a]\widetilde\phi\big\rangle\leq0.
\end{gather*}
We have used the fact that $\Omega$ commutes with every element of $\widetilde{\mathcal{A}}$ as well as with
$\gamma_\mathcal{M}$ and $\mathcal{J}$.
\end{proof}

\section{Two-dimensional f\/lat almost commutative space-time}
\label{Example}

In this section, we construct a~two-dimensional Lorentzian almost commutative geometry as the product of
a~two-dimensional Minkowski space-time and a~noncommutative algebra of complex $2\times 2$ matrices.
This is the simplest example of a~noncommutative space-time, the causal structure of which exhibits new
properties in comparison to a~classical Lorentzian manifold.
We shall present the explicit calculation of the causal structure for pure states as well as for a~class of
mixed states.

The commutative part is constructed as follows:
\begin{itemize}\itemsep=0pt
\item $\H_\mathcal{M} = L^2({\mathbb R}^{1,1}) \otimes {\mathbb C}^{2}$ is the Hilbert space
of square integrable sections of the spinor bundle over the two-dimensional Minkowski space-time.
\item $\mathcal{A}_\mathcal{M} = \mathcal{S}({\mathbb R}^{1,1})$ is the algebra of Schwartz functions
(rapidly decreasing at inf\/inity together with all derivatives) with pointwise multiplication.
\item $\widetilde{\mathcal{A}}_\mathcal{M} = \Span_{{\mathbb C}}(\mathcal{C}_\mathcal{M}) \subset
\mathcal{B}({\mathbb R}^{1,1})$ is a~sub-algebra of the algebra of smooth bounded functions with all derivatives bounded with pointwise multiplication.
{}$\mathcal{C}_\mathcal{M}$ represents the set of smooth bounded causal functions (real-valued functions
non-decreasing along future directed causal curves) on ${\mathbb R}^{1,1}$ with all derivatives bounded.
\item $D_\mathcal{M} = -i \gamma^\mu\partial_\mu$ is the f\/lat Dirac operator.
\item $\mathcal{J}_\mathcal{M}=-[D,x^0] = ic(dx^0) = i\gamma^0$ where $x^0$ is the global time coordinate
and $c$ the Clif\/ford action.
\item $\gamma_\mathcal{M} = \gamma^0 \gamma^1.$
\end{itemize}

This construction is standard for non-unital spectral triples (see~\cite{Gayral} or~\cite{F5}) except that
the unitisation $\widetilde{\mathcal{A}}_\mathcal{M}$ is slightly restricted in order to meet the
condition~\eqref{causality2}.
{}$\widetilde{\mathcal{A}}_\mathcal{M}$ is a~unital algebra which corresponds to a~particular
compactif\/ication of~${\mathbb R}^{1,1}$ \cite{Bes,F6}, hence a~valid unitisation of~$\mathcal{A}_\mathcal{M}$ respecting the axioms.
On compact ordered spaces, monotonic functions which are order-preserving (isotone functions) can span the
entire algebra of functions as a~consequence of the Stone-Weierstrass theorem, whereas on noncompact spaces
such functions must respect some limit condition in order to remain bounded.
In our construction, $\widetilde{\mathcal{A}}_\mathcal{M}$ represents the sub-algebra of functions in
$\mathcal{B}({\mathbb R}^{1,1})$ with existing limits along every causal curve.

The f\/inite and noncommutative part is constructed as a~compact Riemannian spectral triple:
\begin{itemize}\itemsep=0pt
\item $\H_F = {\mathbb C}^2$; \item $\mathcal{A}_F = M_2({\mathbb C})$ is the algebra
of $2\times2$ complex matrices with natural multiplication and representation on $\H_F$;
\item $D_F =
\diag(d_1,d_2)$, with $d_1,d_2 \in {\mathbb R}$.
\end{itemize}

We can restrict our considerations to $D_F$ diagonal since, on the strength of Proposition~\ref{Unitary},
the results are easily translated to any other selfadjoint operator with a~unitary transformation $U \in
M_2({\mathbb C})$.

The f\/inal almost commutative Lorentzian spectral triple is constructed as follows:
\begin{itemize}
\itemsep=0pt \item $\H = L^2({\mathbb R}^{1,1}) \otimes {\mathbb C}^{2} \otimes {\mathbb C}^{2} \cong
L^2({\mathbb R}^{1,1}) \otimes {\mathbb C}^{4}$, \item $\mathcal{A} = \mathcal{S}({\mathbb R}^{1,1})
\otimes M_2({\mathbb C})$, \item $\widetilde{\mathcal{A}} \subset \widetilde{\mathcal{A}}_0 =
\widetilde{\mathcal{A}}_\mathcal{M} \otimes M_2({\mathbb C}) \subset \mathcal{B}({\mathbb R}^{1,1}) \otimes
M_2({\mathbb C})$, \item $D = -i \gamma^\mu\partial_\mu \otimes 1 + \gamma^0 \gamma^1 \otimes
\diag(d_1,d_2) = \diag(-i \gamma^\mu\partial_\mu + \gamma^0 \gamma^1 d_i)_{i\in\{1,2\}}$, \item
$\mathcal{J}= i\gamma^0 \otimes 1 = \diag(i\gamma^0)$.
\end{itemize}

Following the procedure described in Seciton~\ref{General} we f\/irst chose the maximal possible
unitisation $\widetilde{\mathcal{A}}_0 = \Span_{{\mathbb C}}(\mathcal{C}_\mathcal{M}) \otimes M_2({\mathbb
C})$.

Let us now characterise the set $\mathcal{C} = \big\{ \mathbf{a} \in \widetilde{\mathcal{A}}_0: \mathbf{a} =
\mathbf{a}^* \text{ and } \forall\, \phi \in \H,\scal{\phi,\mathcal{J}[D,\mathbf{a}] \phi} \leq 0\big\}$.
For further convenience we shall write an element $\mathbf{a} \in \mathcal{C}$ as $\left(
\begin{smallmatrix}
a & -c
\\
-c^* & b
\end{smallmatrix}
\right)$.
\begin{proposition}
\label{propC}
Let $\mathbf{a} = \left(
\begin{smallmatrix}
a & -c
\\
-c^* & b
\end{smallmatrix}
\right) \in \widetilde{\mathcal{A}}_0$ be a~Hermitian element, then the following conditions are equivalent:
\begin{enumerate}[$(a)$] \itemsep=0pt \item $\mathbf{a}\in\mathcal{C}$, i.e.~$\forall\,\phi \in \H,\scal{\phi,\mathcal{J}[D,\mathbf{a}] \phi}
\leq 0$.
\item At every point of ${\mathbb R}^{1,1}$, the matrix
\begin{gather*}
\left(
\begin{matrix}
a_{,0}+a_{,1} &  0 &  -c_{,0}-c_{,1} &  -(d_1-d_2)c
\\
0 &  a_{,0}-a_{,1} &  (d_1-d_2)c &  -c_{,0}+c_{,1}
\\
-c^*_{,0}-c^*_{,1} &  (d_1-d_2)c^* &  b_{,0}+b_{,1} &  0
\\
-(d_1-d_2)c^* &  -c^*_{,0}+c^*_{,1} &  0 &  b_{,0}-b_{,1}
\end{matrix}
\right)
\end{gather*}
is positive semi-definite.
\item At every point of ${\mathbb R}^{1,1}$, $\forall\, \alpha_1,\alpha_2,\alpha_3,\alpha_4 \in{\mathbb C},$
\begin{gather*}
\abs{\alpha_1}^2(a_{,0}+a_{,1})+\abs{\alpha_2}^2(a_{,0}-a_{,1})+\abs{\alpha_3}^2(b_{,0}+b_{,1}
)+\abs{\alpha_4}^2(b_{,0}-b_{,1})
\\
\qquad
\geq2\Re\left\{\alpha_1^*\alpha_3(c_{,0}+c_{,1})+\alpha_2^*\alpha_4(c_{,0}-c_{,1})\right.
\left.
+\big(\alpha_1^*\alpha_4-\alpha_2^*\alpha_3\big)\abs{d_1-d_2}c\right\}.
\end{gather*}
\end{enumerate}
\end{proposition}

We use here the notation $f_{,\mu} = \partial_\mu f = \pd{f}{x^\mu}$.

\begin{proof}
We work with the following basis of f\/lat gamma matrices:
\begin{gather*}
\gamma^0=\left(
\begin{matrix}
0 &  i
\\
i &  0
\end{matrix}
\right),
\qquad
\gamma^1=\left(
\begin{matrix}
0 &  -i
\\
i &  0
\end{matrix}
\right),
\end{gather*}
which implies
\begin{gather*}
\gamma^0\gamma^1=\gamma_\mathcal{M}=\left(
\begin{matrix}
-1 &  0
\\
0 &  1
\end{matrix}
\right)\cdot
\end{gather*}
Using the isomorphism ${\mathbb C}^{2} \otimes {\mathbb C}^{2} \cong {\mathbb C}^{4}$, we have the
following matrices:
\begin{gather*}
[D,\mathbf{a}]=\left(
\begin{matrix}
-i\gamma^\mu\partial_\mu a &  i\gamma^\mu\partial_\mu c-\gamma_\mathcal{M}(d_1-d_2)c
\\
i\gamma^\mu\partial_\mu c^*+\gamma_\mathcal{M}(d_1-d_2)c^* &  -i\gamma^\mu\partial_\mu b
\end{matrix}
\right),
\qquad
\mathcal{J}=\left(
\begin{matrix}
i\gamma^0 &  0
\\
0 &  i\gamma^0
\end{matrix}
\right),
\\
-\mathcal{J}[D,\mathbf{a}]=\left(
\begin{matrix}
-\gamma^0\gamma^\mu\partial_\mu a &  \gamma^0\gamma^\mu\partial_\mu c-i\gamma^1(d_1-d_2)c
\\
\gamma^0\gamma^\mu\partial_\mu c^*+i\gamma^1(d_1-d_2)c^* &  -\gamma^0\gamma^\mu\partial_\mu b
\end{matrix}
\right)\cdot
\end{gather*}
With our choice of gamma matrices, the above matrix gives precisely the one displayed in~$(b)$.
Now, let us check that the condition $\forall\,\phi \in \H$, $\scal{\phi,\mathcal{J}[D,\mathbf{a}] \phi} \leq 0$
is equivalent to having \mbox{$\mathcal{J}[D,\mathbf{a}] \leq 0$} at every point of ${\mathbb R}^{1,1}$,
i.e.~$-\mathcal{J}[D,\mathbf{a}]$ is a~positive semi-def\/inite matrix at every point of~${\mathbb
R}^{1,1}$.
Indeed, the second condition implies the f\/irst one, and if the second one is false at some particular
point $p\in{\mathbb R}^{1,1}$, then by continuity it is false on some open set $U_p\subset{\mathbb
R}^{1,1}$ and the f\/irst condition is false for some non-null spinor $\phi \in \H$ the support of which is
included in~$U_p$.
Hence, $(a)$~is equivalent to~$(b)$.

The condition $(c)$ is just a~reformulation of the condition $(b)$ with an arbitrary spinor $\phi =
(\alpha_1,\alpha_2,\alpha_3,\alpha_4)$ taken at a~f\/ixed point of ${\mathbb R}^{1,1}$.
\end{proof}

Let us now enlighten the properties of the set $\mathcal{C}$ by considering some of its subsets.
\begin{lemma}
\label{lemmaC}
The following elements in $\widetilde{\mathcal{A}}_0$ belong to $\mathcal{C}$:
\begin{enumerate}[$(a)$]\itemsep=0pt
\item $\mathbf{a} = \left(
\begin{matrix}
a & 0
\\
0 & b
\end{matrix}
\right)$, where $a$ and $b$ are two causal functions on ${\mathbb R}^{1,1}$
($a,b\in\mathcal{C}_\mathcal{M}$).
\item $\mathbf{b} = \left(
\begin{matrix}
a & -c
\\
-c^* & a
\end{matrix}
\right)$, with $a_{,0}-\abs{a_{,1}} \geq \abs{c_{,0}} + \abs{c_{,1}} + \abs{d_1-d_2} \abs{c}$.
\end{enumerate}
\end{lemma}
\begin{proof}
$(a)$  Since a~causal function on $\mathcal{M}$ is non-decreasing along future directed causal curves,
$a,b \in \mathcal{C}_\mathcal{M}$ is equivalent to $a_{,0} \pm a_{,1} \geq 0 $ and $ b_{,0} \pm b_{,1} \geq
0 $, so Proposition~\ref{propC}$(c)$ is respected for $c=0$.
We notice that, by choosing $\alpha_i = 1$ and $\alpha_j = 0$ ($j \neq i$) for $i=1,\dots,4$, this
condition is necessary and suf\/f\/icient whenever $c=0$.

$(b)$ By Proposition~\ref{propC}$(c)$, we must check for all $\alpha_1,\alpha_2,\alpha_3,\alpha_4
\in{\mathbb C}$ the following condition:
\begin{gather*}
\big(\abs{\alpha_1}^2+\abs{\alpha_3}^2\big)(a_{,0}+a_{,1})+\big(\abs{\alpha_2}^2+\abs{\alpha_4}^2\big)(a_{,0}-a_{,1})
\\
\qquad
\geq2\Re\left\{\alpha_1^*\alpha_3(c_{,0}+c_{,1})+\alpha_2^*\alpha_4(c_{,0}-c_{,1})\right.
\left.
+\big(\alpha_1^*\alpha_4-\alpha_2^*\alpha_3\big)\abs{d_1-d_2}c\right\}.
\end{gather*}
Using AM-GM inequality, we have
\begin{gather*}
\big(\abs{\alpha_1}^2+\abs{\alpha_3}^2\big)(a_{,0}+a_{,1})+\big(\abs{\alpha_2}^2+\abs{\alpha_4}^2\big)(a_{,0}-a_{,1})
\\
\qquad
\geq\big(\abs{\alpha_1}^2+\abs{\alpha_3}^2+\abs{\alpha_2}^2+\abs{\alpha_4}^2\big)(\abs{c_{,0}}+\abs{c_{,1}}
+\abs{d_1-d_2}\abs{c})
\\
\qquad
\geq\big(\abs{\alpha_1}^2+\abs{\alpha_3}^2\big)\abs{c_{,0}+c_{,1}}+\big(\abs{\alpha_2}^2+\abs{\alpha_4}^2\big)\abs{c_{,0}-c_{,1}}
\\
\qquad
\phantom{\geq}
{}+\big(\big(\abs{\alpha_1}^2+\abs{\alpha_4}^2\big)+\big(\abs{\alpha_2}^2+\abs{\alpha_3}^2\big)\big)\abs{d_1-d_2}\abs{c}
\\
\qquad
\geq2\abs{\alpha_1}\abs{\alpha_3}\abs{c_{,0}+c_{,1}}+2\abs{\alpha_2}\abs{\alpha_4}\abs{c_{,0}-c_{,1}}
+2(\abs{\alpha_1}\abs{\alpha_4}+\abs{\alpha_2}\abs{\alpha_3})\abs{d_1-d_2}\abs{c})
\\
\qquad
\geq2\Re\left\{\alpha_1^*\alpha_3(c_{,0}+c_{,1})\right\}+2\Re\left\{\alpha_2^*\alpha_4(c_{,0}-c_{,1})\right\}
+2\Re\left\{(\alpha_1^*\alpha_4-\alpha_2^*\alpha_3)\abs{d_1-d_2}c\right\}.
\tag*{\qed}
\end{gather*}
\renewcommand{\qed}{}
\end{proof}
\begin{proposition}
If the unitisation $\widetilde{\mathcal{A}} \subset \widetilde{\mathcal{A}}_0$ is restricted to matrices with
off-diagonal entries
in $\mathcal{A}_\mathcal{M} = \mathcal{S}({\mathbb R}^{1,1}) \subset \Span_{{\mathbb C}}(\mathcal{C}_\mathcal{M})$,
then $\mathcal{C}$ is a~causal cone for the Lorentzian spectral triple
$(\mathcal{A},\widetilde{\mathcal{A}},\H,D,\mathcal{J})$.
\end{proposition}
\begin{proof}
{}$\widetilde{\mathcal{A}}$ is a~unital algebra which contains $\mathcal{A}$ as an ideal.
Let us designate the sub-algebra of Hermitian elements in $\widetilde{\mathcal{A}}$ by
$\widetilde{\mathcal{A}}^H$.
Any element in $\widetilde{\mathcal{A}}^H$ can be written as $\mathbf{a} = \left(
\begin{smallmatrix}
a & -c
\\
-c^* & b
\end{smallmatrix}
\right)$ with $a,b \in \Span_{{\mathbb R}}(\mathcal{C}_\mathcal{M})$ and $c \in \mathcal{S}({\mathbb
R}^{1,1})$.
Then $\mathbf{a}$ can be decomposed as:
\begin{gather*}
\mathbf{a}=\left(
\begin{matrix}
a &  -c
\\
-c^* &  b
\end{matrix}
\right)=\left(
\begin{matrix}a &  0
\\
0 &  b
\end{matrix}
\right)+\left(
\begin{matrix}\widetilde a &  -c
\\
-c^* &  \widetilde a
\end{matrix}
\right)-\left(
\begin{matrix}\widetilde a &  0
\\
0 &  \widetilde a
\end{matrix}
\right),
\end{gather*}
where $\widetilde a$ is a~suitable causal function in $\mathcal{C}_\mathcal{M}$ such that $ \widetilde
a_{,0} \geq \abs{c_{,0}} + \abs{c_{,1}} + \abs{d_1-d_2} \abs{c}$ and $\widetilde a_{,1}=0$.
Such function always exists since $\abs{c_{,0}} + \abs{c_{,1}} + \abs{d_1-d_2} \abs{c}$ is bounded and
rapidly decreasing at inf\/inity.
From Lemma~\ref{lemmaC}, the last two matrices belong to $\mathcal{C}$ and the f\/irst one belongs to
$\Span_{{\mathbb R}}(\mathcal{C})$, so $\mathbf{a} \in \Span_{{\mathbb R}}(\mathcal{C})$ which implies
$\widetilde{\mathcal{A}}^H = \Span_{{\mathbb R}}(\mathcal{C})$.
We conclude by noting that $\widetilde{\mathcal{A}} = \widetilde{\mathcal{A}}^H + i \widetilde{\mathcal{A}}^H =
\Span_{{\mathbb C}}(\mathcal{C})$.
\end{proof}

Hence, we def\/ine the ultimate unitisation $\widetilde{\mathcal{A}}$ as follows:
\begin{gather*}
\widetilde{\mathcal{A}}\cong\left(
\begin{matrix}
\widetilde{\mathcal{A}}_\mathcal{M} &  \mathcal{A}_\mathcal{M}
\\
\mathcal{A}_\mathcal{M} &  \widetilde{\mathcal{A}}_\mathcal{M}
\end{matrix}
\right)=\left(
\begin{matrix}\Span_{{\mathbb C}}(\mathcal{C}_\mathcal{M}) &  \mathcal{S}\big({\mathbb R}^{1,1}\big)
\\
\mathcal{S}\big({\mathbb R}^{1,1}\big) &  \Span_{{\mathbb C}}(\mathcal{C}_\mathcal{M})
\end{matrix}
\right).
\end{gather*}

\begin{figure}[ht] \centering
\includegraphics{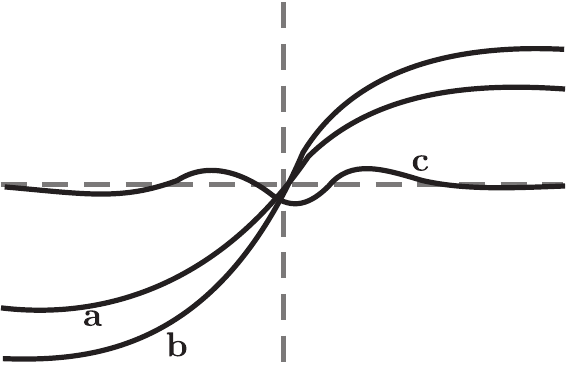}

\caption{Typical behaviour of the functions constituting a~causal element $\mathbf{a} \in
\mathcal{C}$ along a~causal curve.} \label{figcausalcurve}
\end{figure}

Thus we have a~well-def\/ined causal structure on the two-dimensional f\/lat almost commutative space-time
$(\mathcal{A},\widetilde{\mathcal{A}},\H,D,\mathcal{J})$ determined by the set of causal elements
$\mathcal{C}$ characterised in Proposition~\ref{propC}.
The subsets of causal elements def\/ined in Lemma~\ref{lemmaC} give us some information about the general
behaviour of causal elements.
The diagonal entries must be causal functions on ${\mathbb R}^{1,1}$ (i.e.~real and non-decreasing along
causal curves), while the of\/f-diagonal entry is a~complex function which respects a~bound depending on
the growth rate of the diagonal entries along causal curves.
The Fig.~\ref{figcausalcurve} illustrates this behaviour.

\subsection{Causal structure for pure states}
\label{secps}

In this section, we make a~complete characterisation of the causal structure of the two-dimen\-sio\-nal f\/lat
almost commutative space-time $(\mathcal{A},\widetilde{\mathcal{A}},\H,D,\mathcal{J})$ at the level of pure
states.
We shall focus on the space of physical pure states $\mathcal{M}(\widetilde{\mathcal{A}})$, def\/ined
by~\eqref{MA}, that do not localise at inf\/inity.

On the strength of Proposition~\ref{PureStates}
a~pure state $\omega_{p,\xi} \in \mathcal{M}(\widetilde{\mathcal{A}})$ is determined by two entries:
\begin{itemize}\itemsep=0pt
\item a~point $p \in {\mathbb R}^{1,1}$, \item a~normalised complex vector $\xi\in{\mathbb
C}^2$.
\end{itemize}
The value of a~pure state on some element $\mathbf{a}\in\widetilde{\mathcal{A}}$ is:
\begin{gather*}
\omega_{p,\xi}(\mathbf{a})=\xi^*\mathbf{a}(p)\xi,
\end{gather*}
where $\mathbf{a}(p)$ is an application of the evaluation map at $p$ on each entry in $\mathbf{a}$.

It is clear that two normalised vectors determine the same state if and only if they are equal up to
a~phase, so the set of normalised vectors corresponds to ${\mathbb C} P^1 \cong S^2$.
Hence, on the set-theoretic level, the space of pure states $\mathcal{M}(\widetilde{\mathcal{A}})$ is
isomorphic to the product ${\mathbb R}^{1,1} \times S^2$.
We will refer to ${\mathbb R}^{1,1}$ as the continuous space and to $S^2$ as the internal space.
Hoverer, one should be aware that the geometry of the internal space, resulting from the f\/inite
noncommutative algebra, is quite dif\/ferent from the usual Riemannian geometry of $S^2$
(see~\cite{IochKM}).
\begin{proposition}
\label{propcondcausal}
Let us take two pure states $\omega_{p,\xi},\omega_{q,\varphi} \in \mathcal{M}(\widetilde{\mathcal{A}})$.
Then $\omega_{p,\xi} \preceq \omega_{q,\varphi}$ if and only if
$\forall\, \mathbf{a} = \left(\begin{smallmatrix} a & -c\\-c^* & b\end{smallmatrix}\right) \in \mathcal{C}$,
\begin{gather*}
\abs{\varphi_1}^2a(q)-\abs{\xi_1}^2a(p)+\abs{\varphi_2}^2b(q)-\abs{\xi_2}
^2b(p)\geq2\Re\big\{\varphi^*_1\varphi_2c(q)-\xi^*_1\xi_2c(p)\big\},
\end{gather*}
where $\xi = \Big(\!\begin{smallmatrix} \xi_1\\\xi_2\end{smallmatrix}\!\Big)$
and $\varphi =  \Big(\!\begin{smallmatrix}\varphi_1\vspace{1mm}\\\varphi_2\end{smallmatrix}\!\Big)$.
\end{proposition}
\begin{proof}
From Def\/inition~\ref{defcausal}, $\omega_{p,\xi} \preceq \omega_{q,\varphi}$ if and only if $\forall\,
\mathbf{a}\in\mathcal{C}$,
\begin{gather*}
\omega_{p,\xi}(\mathbf{a})\leq\omega_{q,\varphi}(\mathbf{a})
\quad
\Longleftrightarrow
\quad
\xi^*\mathbf{a}
(p)\xi\leq\varphi^*\mathbf{a}(q)\varphi
\\
\Longleftrightarrow
\quad
\left(
\begin{matrix}
\xi^*_1 &  \xi^*_2
\end{matrix}
\right)\left(
\begin{matrix}
a(p) &  -c(p)
\\
-c(p)^* &  b(p)
\end{matrix}
\right)\left(
\begin{matrix}
\xi_1
\\
\xi_2
\end{matrix}
\right) \leq \left(
\begin{matrix}
\varphi^*_1 &  \varphi^*_2
\end{matrix}
\right)\left(
\begin{matrix}
a(q) &  -c(q)
\\
-c(q)^* &  b(q)
\end{matrix}
\right)\left(
\begin{matrix}
\varphi_1
\\
\varphi_2
\end{matrix}
\right)
\\
\Longleftrightarrow
\quad
\abs{\xi_1}^2a(p)-2\Re\big\{\xi^*_1\xi_2c(p)\big\}+\abs{\xi_2}^2b(p)\leq\abs{\varphi_1}
^2a(q)-2\Re\big\{\varphi^*_1\varphi_2c(q)\big\}+\abs{\varphi_2}^2b(q).\!\!\!\!\tag*{\qed}
\end{gather*}
\renewcommand{\qed}{}
\end{proof}

\begin{proposition}
\label{propcaus1}
Two pure states $\omega_{p,\xi},\omega_{q,\varphi} \in \mathcal{M}(\widetilde{\mathcal{A}})$ are causally
related with $\omega_{p,\xi} \preceq \omega_{q,\varphi}$ only if $p \preceq q$ in ${\mathbb R}^{1,1}$.
\end{proposition}
\begin{proof}
From Lemma~\ref{lemmaC}, $\mathbf{a} = \left(
\begin{smallmatrix}
a & 0
\\
0 & a
\end{smallmatrix}
\right) \in \mathcal{C}$ where $a$ is a~causal function on ${\mathbb R}^{1,1}$.
Using Proposition~\ref{propcondcausal} with $\mathbf{a}$
we f\/ind $\big(\abs{\varphi_1}^2+\abs{\varphi_2}^2\big)$ $a(q)-\big(\abs{\xi_1}^2 + \abs{\xi_2}^2\big)$ $a(p) \geq 0 \implies a(p) \leq a(q)$.
Since the above inequality is valid for all $a \in \mathcal{C}_\mathcal{M}$, we have $p \preceq q$ from
Theorem~\ref{thmreconstruction}.
\end{proof}

This f\/irst necessary condition tells us that the coupling of ${\mathbb R}^{1,1}$ with the internal space
$S^2$ does not induce any violation on the usual causal structure on ${\mathbb R}^{1,1}$ and the result
does not depend on the particular form of $D_F$.
Let us now show that if the f\/inite-part Dirac operator is degenerate (i.e.~$d_1 = d_2$), then no changes
in the internal space are allowed by the causal structure.
\begin{proposition}
\label{Dfdeg}
If the finite-part Dirac operator is degenerate $(d_1 = d_2)$, then two pure states
$\omega_{p,\xi},\omega_{q,\varphi} \in \mathcal{M}(\widetilde{\mathcal{A}})$ are causally related with
$\omega_{p,\xi} \preceq \omega_{q,\varphi}$ only if $p \preceq q$ in ${\mathbb R}^{1,1}$ and $\xi =\varphi$.
\end{proposition}

\begin{proof}
If $d_1 = d_2$, then the element $\mathbf{a} = \left(
\begin{smallmatrix}
a & -c
\\
-c^* & b
\end{smallmatrix}
\right) \in \widetilde{\mathcal{A}}$, with $a, b$ and $c$ being constant functions, is in $\mathcal{C}$ on
the strength of Proposition~\ref{propC}.
For such element $\mathbf{a}$ the causality condition form Proposition~\ref{propcondcausal} reads
\begin{gather*}
\big(\abs{\varphi_1}^2-\abs{\xi_1}^2\big)a+\big(\abs{\varphi_2}^2-\abs{\xi_2}^2\big)b\geq2\Re\big\{(\varphi^*_1\varphi_2-\xi^*_1\xi_2)c\big\}.
\end{gather*}
Since $a,b \in {\mathbb R}$ and $c \in {\mathbb C}$ are arbitrary the above inequality is always
fulf\/illed only if $\xi = \varphi$.
\end{proof}

From now on we will assume that the f\/inite-part Dirac operator is not degenerate (i.e.\ $d_1 \neq d_2$).
Under this assumption, the movements in the internal space are allowed, but they turn out to be further
constrained by the causal structure.
In order to express the second necessary condition it is convenient to write down explicitly the
isomorphism between~${\mathbb C} P^1$ and~$S^2$.
A~normalised vector $\xi=(\xi_1,\xi_2) \in {\mathbb C} P^1$ can be associated with the point
$(x_\xi,y_\xi,z_\xi)$ of~$S^2$ embedded in ${\mathbb R}^3$ by the relations:
\begin{gather*}
x_\xi=2\Re\{\xi_1^*\xi_2\},
\qquad
y_\xi=2\Im\{\xi_1^*\xi_2\},
\qquad
z_\xi=\abs{\xi_1}^2-\abs{\xi_2}^2.
\end{gather*}
We will refer to $z_\xi$ as the latitude (or the altitude) of the vector $\xi$ on $S^2$.
\begin{proposition}
\label{propcaus2}
Two pure states $\omega_{p,\xi},\omega_{q,\varphi} \in \mathcal{M}(\widetilde{\mathcal{A}})$ are causally
related with $\omega_{p,\xi} \preceq \omega_{q,\varphi}$ only if~$\xi$ and~$\varphi$ have the same
latitude, i.e.~$z_\xi=z_\varphi$.
\end{proposition}
\begin{proof}
From Lemma~\ref{lemmaC}, $\mathbf{a} = \left(
\begin{smallmatrix}
a & 0
\\
0 & 0
\end{smallmatrix}
\right) \in \mathcal{C}$ where $a$ is a~constant function on ${\mathbb R}^{1,1}$.
Using Proposition~\ref{propcondcausal} with $\mathbf{a}$ we f\/ind:
\begin{gather*}
\abs{\varphi_1}^2a(q)-\abs{\xi_1}^2a(p)=\big(\abs{\varphi_1}^2-\abs{\xi_1}^2\big)a\geq0.
\end{gather*}
Since this inequality must be valid for all $a \in {\mathbb R}$, we must have $\abs{\varphi_1}^2 =
\abs{\xi_1}^2$.
Since the vectors are normalised we must have $\abs{\varphi_2}^2 = \abs{\xi_2}^2$, hence $z_\xi=z_\varphi$.
\end{proof}

This condition is completely coherent with the results obtained in~\cite{IochKM} using Connes' Riemannian
distance formula on the noncommutative space $M_2({\mathbb C})$.
The parallels of latitude within the sphere $S^2$ were found to be separated by an inf\/inite distance.
This can be viewed as a~geometrical characteristic of our internal space, which forbids any movement
between disconnected parallels.
From Proposition~\ref{Unitary}, it follows that a~unitary transformation of the f\/inite-part Dirac
operator $D_F$ is equivalent to a~rotation in the internal space of states.
Hence, the notion of the latitude is completely determined by the diagonalisation of the operator $D_F$.

We should remark that the inf\/inite distance separation between the parallels is a~property of the
representation of $M_2({\mathbb C})$ on ${\mathbb C}^2$, which is the usual type of representation for the
noncommutative standard model~\cite{CCM06,Dungen}.
It has been shown in~\cite{CDMW} that one can obtain a~f\/inite distance between the parallels by using
instead a~representation on~$M_2({\mathbb C})$ and by def\/ining an opera\-tor~$D$ as a~truncation of the
usual Dirac operator on the Moyal plane.
It is also not known if the correspondence between causally disconnected states and inf\/inite Riemannian
distance is just a~coincidence of this particular model or a~more general property.

We now come to our main result which is the complete characterisation of the motions in the internal space
that do not violate the causality conditions.
It turns out that the possible movements within a~particular parallel of f\/ixed latitude in the internal
space are intrinsically connected to the motion in the space-time.

In order to simplify the expressions, we can assume without loss of generality that $\xi_1$ and $\varphi_1$
are always real and non-negative since $\xi$ and $\varphi$ are def\/ined up to phase.
Since $\xi$ and $\varphi$ have the same latitude and are normalised vectors, we f\/ind that
$\xi_1=\varphi_1$ and $\abs{\xi_2}=\abs{\varphi_2}$, so we can set $\xi_2 = \abs{\varphi_2}
e^{i\theta_\xi}$ and $\varphi_2 = \abs{\varphi_2} e^{i\theta_\varphi}$ with $\theta_\xi$ and
$\theta_\varphi$ representing the position of the vectors on the parallel of latitude.
\begin{theorem}
\label{thmcausal}
Two pure states $\omega_{p,\xi},\omega_{q,\varphi} \in \mathcal{M}(\widetilde{\mathcal{A}})$ are causally
related with $\omega_{p,\xi} \preceq \omega_{q,\varphi}$ if and only if the following conditions are
respected:
\begin{itemize}\itemsep=0pt
\item $p \preceq q$ in ${\mathbb R}^{1,1}$; \item $\xi$ and $\varphi$ have the same latitude;
\item $l(\gamma) \geq \frac{\abs{\theta_\varphi-\theta_\xi}}{\abs{d_1-d_2}}$, where $l(\gamma)$ represents
the length of a~causal curve $\gamma$ going from $p$ to $q$ on~${\mathbb R}^{1,1}$.
\end{itemize}
\end{theorem}

The proof of this theorem will be completed in the remaining part of this section.
The necessary parts of the two f\/irst conditions have already been treated in Propositions~\ref{propcaus1}
and~\ref{propcaus2}.
The suf\/f\/icient conditions are proven in Proposition~\ref{propcaus3} while the necessary condition on
$l(\gamma)$ is proven in Proposition~\ref{propcaus4}.
\begin{figure}[ht] \centering
\includegraphics{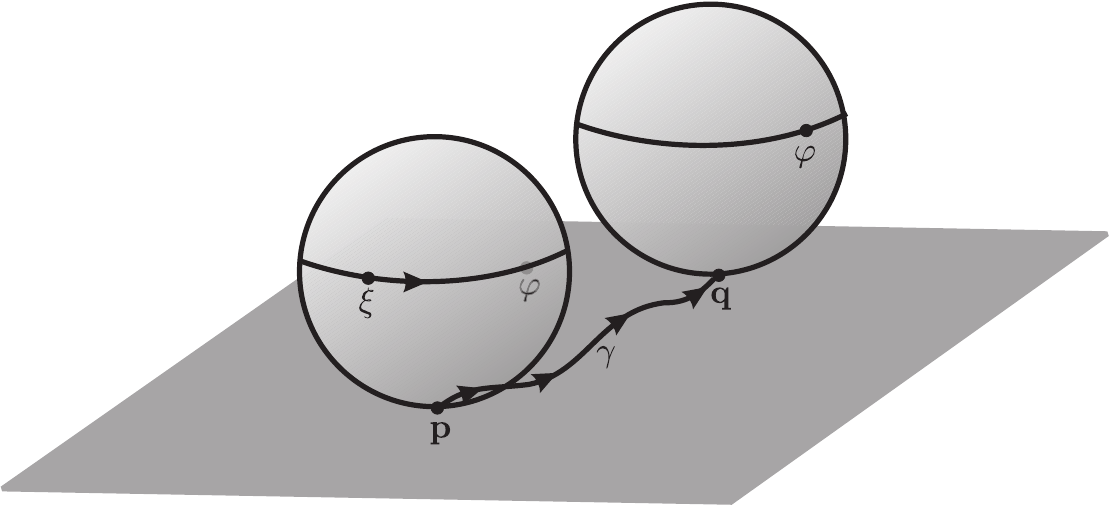}
\caption{Representation of a~path in the space of pure states.}
\label{figcausalrel}
\end{figure}

Theorem~\ref{thmcausal} gives a~full description of the causal structure of the two-dimensional f\/lat
almost commutative space-time.
A~graphical presentation of the space of pure states with a~visualisation of a~causal path is given in
Fig.~\ref{figcausalrel}.
Whenever the curve $\gamma$ is not null, the inequality between the distances in the internal and
continuous spaces can be written as:
\begin{gather}
\label{speedoflight}
\frac{\abs{\theta_\varphi-\theta_\xi}}{l(\gamma)}\leq\abs{d_1-d_2}.
\end{gather}

{}$\abs{\theta_\varphi-\theta_\xi}$ represents a~measure of the distance within the parallel of latitude
between two vectors in the internal space.
{}$l(\gamma) = \int \sqrt{-g_\gamma(\dot \gamma(t),\dot \gamma(t))} {\rm  d}t $ physically represents the proper
time of $\gamma$, i.e.~the time measured by a~clock moving along $\gamma$ in the Minkowski space-time.
{}$\abs{d_1-d_2}$ is a~constant def\/ined by the eigenvalues of the operator $D_F$ of the internal space.
The condition~\eqref{speedoflight} can be seen as a~constant upper bound to the ratio between the distance
in the internal space and the proper time in the continuous space.
Note that if one moves along a~light-like geodesic in the Minkowski space-time (i.e.~$l(\gamma) = 0$), then
no motion in the internal space is allowed by the causal structure.

We now come to the technical proof of Theorem~\ref{thmcausal}.
At f\/irst, we consider a~timelike curve $\gamma$ linking the space-time events $p$ and $q$.
The following Proposition will give us the suf\/f\/icient part of Theorem~\ref{thmcausal}.
\begin{proposition}
\label{propcaus3}
Suppose that $p \preceq q$ with $\gamma$ a~future directed timelike curve going from $p$ to $q$ and that
$\xi$, $\varphi$ have the same latitude, with $\xi_1=\varphi_1\in{\mathbb R}^*$ and $\xi_2 =
\abs{\varphi_2} e^{i\theta_\xi}$, $\varphi_2 = \abs{\varphi_2} e^{i\theta_\varphi}$.
If $l(\gamma) \geq \frac{\abs{\theta_\varphi-\theta_\xi}}{\abs{d_1-d_2}}$, then $\omega_{p,\xi} \preceq
\omega_{q,\varphi}$.
\end{proposition}
\begin{proof}

Since $\varphi_1 = \xi_1\in{\mathbb R}$ and $\abs{\varphi_2} = \abs{\xi_2}$, from
Proposition~\ref{propcondcausal} we need to prove the following inequality $\forall\, \mathbf{a} = \left(
\begin{smallmatrix}
a & -c
\\
-c^* & b
\end{smallmatrix}
\right) \in \mathcal{C}$,
\begin{gather}
\label{goal}
\abs{\varphi_1}^2(a(q)-a(p))+\abs{\varphi_2}
^2(b(q)-b(p))\geq2\Re\big\{\varphi_1\varphi_2c(q)-\varphi_1\xi_2c(p)\big\}.
\end{gather}

Let us f\/irst take a~timelike curve $\gamma: [0,T] \to {\mathbb R}^{1,1}$, with $\gamma(0)=p$,
$\gamma(T)=q$ and $l(\gamma) = \frac{\abs{\theta_\varphi-\theta_\xi}}{d}$, where $d=\abs{d_1-d_2}$ and $T
>0$.
We def\/ine the following functions for $t\in[0,T]$:
\begin{itemize}
\itemsep=0pt \item $v(t)= \frac{\dot\gamma_1(t)}{\dot\gamma_0(t)} \in ]{-}1,1[$ with $\dot\gamma(t) =
(\dot\gamma_0(t),\dot\gamma_1(t))$, which is well def\/ined since $\gamma$ is future directed timelike, so
$\dot\gamma_0(t) > 0$ and $\dot\gamma_0(t) > \abs{\dot\gamma_1(t)}$.
\item $\lambda_1(t) = \frac{1 + v(t)}{2}$, $\lambda_2(t) = \frac{1-v(t)}{2} \in]0,1[$.
\item $\chi(t) = \xi_2  e^{i \sgn(\theta_\varphi-\theta_\xi)  d  l_\gamma(t)}$ where
$l_\gamma(t)=\int_0^t \sqrt{-g_{\gamma(s)}(\dot \gamma(s),\dot \gamma(s))}  ds$ represents
the length of~$\gamma$ restricted to the interval $[0,t]$, with $t \leq T$.
We have obviously that $\abs{\chi(t)}= \abs{\xi_2}=\abs{\varphi_2} $ $\forall\, t\in[0,T]$, $\chi(0)=\xi_2$
and $\chi(T)=\varphi_2$.
\end{itemize}

Using the second fundamental theorem of calculus, we have the following equality:
\begin{gather*}
a(q)-a(p)=a(\gamma(T))-a(\gamma(0))
\\
\hphantom{a(q)-a(p)}{} =\int_0^T\mathrm{d}a_{\gamma(t)}(\dot\gamma(t)) dt
=\int_0^T\dot\gamma_0(t)\prt{a_{,0}(\gamma(t))+v(t) a_{,1}(\gamma(t))} dt
\\
\hphantom{a(q)-a(p)}{} =\int_0^T\dot\gamma_0(t)\left[\lambda_1(t)\prt{a_{,0}(\gamma(t))+a_{,1}(\gamma(t))}
+\lambda_2(t)\prt{a_{,0}(\gamma(t))-a_{,1}(\gamma(t))}\right] dt.
\end{gather*}
In the following, we shall omit the arguments $t$ and $\gamma(t)$ in order to have more readable
expressions.
An analogous equality can be obtained for the function $b$:
\begin{gather*}
b(q)-b(p)=\int_0^T\dot\gamma_0\left[\lambda_1(b_{,0}+b_{,1})+\lambda_2(b_{,0}-b_{,1})\right] dt.
\end{gather*}
Now, the l.h.s.\ of~\eqref{goal} can be written as:
\begin{gather}
\abs{\varphi_1}^2(a(q)-a(p))+\abs{\varphi_2}^2(b(q)-b(p))
\nonumber
\\
\qquad{}
=\int_0^T\dot\gamma_0\left[\abs{\varphi_1}^2(\lambda_1(a_{,0}+a_{,1})+\lambda_2(a_{,0}-a_{,1}
))+\abs{\varphi_2}^2(\lambda_1(b_{,0}+b_{,1})+\lambda_2(b_{,0}-b_{,1}))\right]dt
\nonumber
\\
\qquad{}
=\int_0^T\dot\gamma_0\Big[\abs{\varphi_1}^2\lambda_1(a_{,0}+a_{,1})+\abs{\varphi_1}^2\lambda_2(a_{,0}
-a_{,1})
\nonumber
\\
\qquad\phantom{=}
{}+\abs{\chi}^2\lambda_1(b_{,0}+b_{,1})+\abs{\chi}^2\lambda_2(b_{,0}-b_{,1})\Big]dt.
\label{insideb}
\end{gather}
If we apply Proposition~\ref{propC}$(c)$ to the integrand of~\eqref{insideb} with
\begin{gather*}
\alpha_1=\sqrt{\lambda_1}\varphi_1e^{i\Delta_1},
\qquad
\alpha_2=\sqrt{\lambda_2}\varphi_1e^{i\Delta_2},
\qquad
\alpha_3=\sqrt{\lambda_1}\chi e^{i\Delta_1},
\qquad
\alpha_4=\sqrt{\lambda_2}\chi e^{i\Delta_2}
\end{gather*}
we get
\begin{gather}
\abs{\varphi_1}^2(a(q)-a(p))+\abs{\varphi_2}^2(b(q)-b(p))
\nonumber
\\
\qquad
\geq\int_0^T\dot\gamma_0\left[2\Re\big\{\varphi_1\chi\lambda_1(c_{,0}+c_{,1})+\varphi_1\chi\lambda_2(c_{,0}
-c_{,1})\big\}\phantom{\sqrt{\lambda_1}}\right.
\nonumber
\\
\qquad\phantom{=}
\left.
{}+2\Re\big\{\sqrt{\lambda_1\lambda_2}\big(e^{i(\Delta_2-\Delta_1)}-e^{-i(\Delta_2-\Delta_1)}
\big)\varphi_1\chi dc\big\}\right]dt
\nonumber
\\
\qquad
=\int_0^T\left[2\Re\left\{\varphi_1\chi\dv{c}{t}\right\}
+2\Re\big\{\varphi_1c \chi i\sin(\Delta_2-\Delta_1) d 2\dot\gamma_0\sqrt{\lambda_1\lambda_2}\big\}\right]dt.
\label{tosubst}
\end{gather}

Now, let us note that:
\begin{gather*}
2\dot\gamma_0(t)\sqrt{\lambda_1(t)\lambda_2(t)}=\dot\gamma_0(t)\sqrt{1-v^2(t)}
=\sqrt{\dot\gamma^2_0(t)-\dot\gamma^2_1(t)}=\sqrt{-g_{\gamma(t)}(\dot\gamma(t),\dot\gamma(t))}
\end{gather*}
and
\begin{gather}
\dv{\chi(t)}{t}=\chi(t) i\sgn(\theta_\varphi-\theta_\xi) d \dv{l_\gamma(t)}{t}
=\chi(t) i\sgn(\theta_\varphi-\theta_\xi) d \sqrt{-g_{\gamma(t)}(\dot\gamma(t),\dot\gamma(t))}
\nonumber
\\
\phantom{\dv{\chi(t)}{t}}
=\chi(t) i\sgn(\theta_\varphi-\theta_\xi) d 2\dot\gamma_0(t)\sqrt{\lambda_1(t)\lambda_2(t)}.
\label{fromsubst}
\end{gather}

So if we choose $\Delta_1$, $\Delta_2$ to have $\sin(\Delta_2-\Delta_1)=\sgn(\theta_\varphi-\theta_\xi)$,
we can substitute~\eqref{fromsubst} into~\eqref{tosubst} to get:
\begin{gather}
\abs{\varphi_1}^2(a(q)-a(p))+\abs{\varphi_2}^2(b(q)-b(p))
\geq\int_0^T\left[2\Re\left\{\varphi_1\chi\dv{c}{t}\right\}+2\Re\left\{\varphi_1c \dv{\chi}{t}\right\}\right]dt
\nonumber
\\
\qquad =\int_0^T2\Re\left\{\varphi_1\dv{}{t}\prt{\chi c}\right\}dt
 =2\Re\big\{\varphi_1 \chi(T) c(\gamma(T))-\varphi_1 \chi(0) c(\gamma(0))\big\} \nonumber
\\
\qquad =2\Re\big\{\varphi_1\varphi_2c(q)-\varphi_1\xi_2c(p)\big\}\label{goto}
\end{gather}
and we have proven~\eqref{goal}.

In order to complete the proof, we must show that this result is still valid when $l(\gamma) \geq
\frac{\abs{\theta_\varphi-\theta_\xi}}{d}$.
Since the causal order is a~well def\/ined partial order, it is suf\/f\/icient to prove that, if
$\xi=\varphi$, then the causal relation is valid for every causal curve $\gamma$ with length $l(\gamma)
\geq 0$.
This can be done by using~\eqref{tosubst} with $\chi$ constant and $\Delta_2=\Delta_1$:
\begin{gather*}%\label{RHSec}
\abs{\varphi_1}^2(a(q)-a(p))+\abs{\varphi_2}^2(b(q)-b(p))
\\
\qquad{}\geq\int_0^T\left[2\Re\left\{\varphi_1\chi\dv{c}{t}\right\}+0\right]dt
=2\Re\big\{\varphi_1\varphi_2c(q)-\varphi_1\xi_2c(p)\big\}.
\tag*{\qed}
\end{gather*}
\renewcommand{\qed}{}
\end{proof}
\begin{proposition}
\label{propcaus4}
Let us suppose that $p \preceq q$ and that $\xi$, $\varphi$ have the same latitude, with
$\xi_1=\varphi_1\in{\mathbb R}^*$ and $\xi_2 = \abs{\varphi_2} e^{i\theta_\xi}$, $\varphi_2 =
\abs{\varphi_2} e^{i\theta_\varphi}$, with $0<\abs{\theta_\varphi-\theta_\xi}\leq\pi$.
If $\omega_{p,\xi} \preceq \omega_{q,\varphi}$, then there exists a~future directed timelike curve $\gamma$
going from $p$ to $q$ such that $l(\gamma) \geq \frac{\abs{\theta_\varphi-\theta_\xi}}{\abs{d_1-d_2}}$.
\end{proposition}
\begin{proof}
At f\/irst, we shall assume that $\abs{\theta_\varphi-\theta_\xi}<\pi$.
Let us note, that we can exclude the cases where $\varphi_1=0$ or $\varphi_2=0$ since they correspond to
the poles of the sphere $S^2$ where no movement is possible within the internal space, so every future
directed timelike curve would be convenient.

We will prove this Proposition by contradiction.
Let us suppose that for every future directed timelike curve $\gamma$ such that $\gamma(0)=p$ and
$\gamma(T)=q$ we have $l(\gamma) < \frac{\abs{\theta_\varphi-\theta_\xi}}{d}$ with $d=\abs{d_1-d_2}$.

In the same way as in the proof of Proposition~\ref{propcaus3}, we def\/ine $\lambda_1(t) = \frac{1 +
v(t)}{2}$ and $\lambda_2(t) = \frac{1-v(t)}{2}$ with $v(t)= \frac{\dot\gamma_1(t)}{\dot\gamma_0(t)}$ for
$t\in[0,T]$.

By Proposition~\ref{propcondcausal}, in order to demonstrate that $\omega_{p,\xi} \preceq
\omega_{q,\varphi}$ is false we only need to f\/ind one causal element $\mathbf{a} = \left(
\begin{smallmatrix}
a & -c
\\
-c^* & b
\end{smallmatrix}
\right) \in \mathcal{C}$ such that:
\begin{gather}
\label{antigoal}
\abs{\varphi_1}^2(a(q)-a(p))+\abs{\varphi_2}
^2(b(q)-b(p))<2\Re\big\{\varphi_1\varphi_2c(q)-\varphi_1\xi_2c(p)\big\}.
\end{gather}

We shall def\/ine such an element along the curve $\gamma$ and consider its implicit extension to the whole
space ${\mathbb R}^{1,1}$.
We f\/irst def\/ine the of\/f-diagonal terms as:
\begin{gather*}
c(\gamma(t))=-\csc(d l_\gamma(t)+\epsilon)e^{i(\sgn(\theta_\varphi-\theta_\xi)\epsilon-\theta_\xi)},
\end{gather*}
where $l_\gamma(t)=\int_0^t\!\sqrt{-g_{\gamma(s)}(\dot \gamma(s),\dot \gamma(s))}  ds$
and $\epsilon > 0$ is chosen such that $\abs{\theta_\varphi-\theta_\xi} + \epsilon<\pi$.
Since $d  l_\gamma(T) < \abs{\theta_\varphi-\theta_\xi}$, this implies that $c(\gamma(t)) $ is well
def\/ined $\forall\, t \in [0,T]$.

The diagonal terms $a$ and $b$ are def\/ined up to a~constant along the curve $\gamma$ as follows:
\begin{gather}
a_{,0}(\gamma(t))=\frac{d}{2\sqrt{\lambda_1(t)\lambda_2(t)}}\frac{\abs{\varphi_2}}{\abs{\varphi_1}}
\csc^2(d l_\gamma(t)+\epsilon),
\nonumber
\\
a_{,1}(\gamma(t))=\frac{-v(t) d}{2\sqrt{\lambda_1(t)\lambda_2(t)}}\frac{\abs{\varphi_2}}{\abs{\varphi_1}}
\csc^2(d l_\gamma(t)+\epsilon),
\nonumber
\\
b_{,0}(\gamma(t))=\frac{d}{2\sqrt{\lambda_1(t)\lambda_2(t)}}\frac{\abs{\varphi_1}}{\abs{\varphi_2}}
\csc^2(d l_\gamma(t)+\epsilon),
\nonumber
\\
b_{,1}(\gamma(t))=\frac{-v(t) d}{2\sqrt{\lambda_1(t)\lambda_2(t)}}\frac{\abs{\varphi_1}}{\abs{\varphi_2}}
\csc^2(d l_\gamma(t)+\epsilon).
\label{diagel}
\end{gather}

The proof of $\mathbf{a}$ being positive semi-def\/inite along the curve $\gamma$ is a~straightforward
algebraic computation and we shift to the Appendix.
Next, we extend $a$, $b$ and $c$ to the whole ${\mathbb R}^{1,1}$ in such a~way that the matrix
$\mathbf{a}$ remains positive semi-def\/inite at each point of the Minkowski space-time.
Such an extension is always possible since $a$, $b$ and $c$ were only def\/ined on a~compact set, so
$c(\gamma)$ can be extended to $c\in \mathcal{S}({\mathbb R}^{1,1})$ and the extension of $a$ and $b$ can
be constructed using e.g.\
the inequalities in Lemma~\ref{lemmaC}.
Since such an extension has no impact on the following arguments, we do not need to specify it explicitly.

By making use of the proof of Proposition~\ref{propcaus3} we obtain the following equality:
\begin{gather*}
\abs{\varphi_1}^2(a(q)-a(p))+\abs{\varphi_2}^2(b(q)-b(p))
\\
\qquad
=\int_0^T\dot\gamma_0\Big[\abs{\varphi_1}^2\lambda_1(a_{,0}+a_{,1})+\abs{\varphi_1}^2\lambda_2(a_{,0}
-a_{,1})\\
\qquad\hphantom{+}{}
+\abs{\varphi_2}^2\lambda_1(b_{,0}+b_{,1})+\abs{\varphi_2}^2\lambda_2(b_{,0}-b_{,1})\Big]dt
\\
\qquad
=\int_0^T\dot\gamma_0\left[4\sqrt{\lambda_1(t)\lambda_2(t)} d \abs{\varphi_1\varphi_2}
\csc^2(d l_\gamma(t)+\epsilon)\right]dt
\\
\qquad
=2\abs{\varphi_1\varphi_2}\int_0^T d 2\dot\gamma_0\sqrt{\lambda_1(t)\lambda_2(t)}
\csc^2(d l_\gamma(t)+\epsilon) dt.
\end{gather*}
Since $2 \dot\gamma_0 \sqrt{\lambda_1(t) \lambda_2(t)} = \sqrt{\dot\gamma^2_0(t)-\dot\gamma^2_1(t)} =
\dv{l_\gamma(t)}{t}$, we can compute the integral explicitly:
\begin{gather}
\label{expr1}
\abs{\varphi_1}^2(a(q)-a(p))+\abs{\varphi_2}^2(b(q)-b(p))=2\abs{\varphi_1\varphi_2}
\prt{-\cot(d l_\gamma(T)+\epsilon)+\cot(\epsilon)}.
\end{gather}

Now we can also make explicit the r.h.s.\ of~\eqref{antigoal}:
\begin{gather}
2\Re\big\{\varphi_1\varphi_2c(q)-\varphi_1\xi_2c(p)\big\}
=2\abs{\varphi_1\varphi_2}\Re\big\{e^{i\theta_\varphi}c(\gamma(T))-e^{i\theta_\xi}c(\gamma(0))\big\}
\nonumber
\\
\qquad
=2\abs{\varphi_1\varphi_2}
\prt{-\Re\big\{e^{i(\theta_\varphi-\theta_\xi+\sgn(\theta_\varphi-\theta_\xi)\epsilon)}
\csc(d l_\gamma(T)+\epsilon)\big\}+\Re\big\{e^{i\sgn(\theta_\varphi-\theta_\xi)\epsilon}\csc(\epsilon)\big\}}
\nonumber
\\
\qquad
=2\abs{\varphi_1\varphi_2}
\prt{-\frac{\cos(\theta_\varphi-\theta_\xi+\sgn(\theta_\varphi-\theta_\xi)\epsilon)}
{\sin(d l_\gamma(T)+\epsilon)}+\frac{\cos(\sgn(\theta_\varphi-\theta_\xi)\epsilon)}{\sin(\epsilon)}}
\nonumber
\\
\qquad
=2\abs{\varphi_1\varphi_2}\prt{-\frac{\cos(\abs{\theta_\varphi-\theta_\xi}+\epsilon)}
{\sin(d l_\gamma(T)+\epsilon)}+\cot(\epsilon)}.
\label{expr2}
\end{gather}

Since the cosine function is strictly decreasing on $]0,\pi[$ and since from our assumption we have $0 <
d l_\gamma(T) + \epsilon < \abs{\theta_\varphi -\theta_\xi} + \epsilon < \pi$, we get the following strict
inequality:
\begin{gather*}
\phantom{\implies}\quad \ \cos(d l_\gamma(T)+\epsilon)>\cos(\abs{\theta_\varphi-\theta_\xi}+\epsilon)
\\
\implies\quad \cot(d l_\gamma(T)+\epsilon)>\frac{\cos(\abs{\theta_\varphi-\theta_\xi}+\epsilon)}
{\sin(d l_\gamma(T)+\epsilon)}
\\
\implies\quad 2\abs{\varphi_1\varphi_2}\prt{-\cot(d l_\gamma(T)+\epsilon)+\cot(\epsilon)}
<2\abs{\varphi_1\varphi_2}\prt{-\frac{\cos(\abs{\theta_\varphi-\theta_\xi}+\epsilon)}
{\sin(d l_\gamma(T)+\epsilon)}+\cot(\epsilon)}
\\
\implies\quad \abs{\varphi_1}^2(a(q)-a(p))+\abs{\varphi_2}
^2(b(q)-b(p))<2\Re\big\{\varphi_1\varphi_2c(q)-\varphi_1\xi_2c(p)\big\}
\end{gather*}
by using the equalities~\eqref{expr1} and~\eqref{expr2}.
Hence~\eqref{antigoal} is true and the proof is complete for $\abs{\theta_\varphi\!-\!\theta_\xi}\!<\!\pi$.

The case $\theta_\varphi = \theta_\xi + \pi$ can be dealt with by using the transitivity of the causal order.
Once more we suppose by contradiction that $d l_\gamma(T) = \pi-\delta < \pi$.
We can construct an extension of the curve~$\gamma$ to $[0,T']$ with $T'>T$ and $q \preceq q'=\gamma(T')$
such that $d l_\gamma(T') = \pi-\frac{2}{3}\delta$.
From Proposition~\ref{propcaus3}, the causal relation $\omega_{q,\varphi}\preceq \omega_{q',\varphi'}$ for
$\theta_{\varphi'} = \theta_{\varphi- \frac 13 \delta}$ is allowed since the length of~$\gamma$ between~$q$
and $q'$ is~$\frac 13 \delta$.
Since $\abs{\theta_{\varphi'}-\theta_\xi} = \pi-\frac 13 \delta <\pi$, we can apply the previous result
to get the~necessary condition $d l_\gamma(T') \geq \pi-\frac 13 \delta$ which contradicts
$d l_\gamma(T') = \pi-\frac{2}{3}\delta$.
Hence $d l_\gamma(T) \geq \pi$.
\end{proof}

The Propositions~\ref{propcaus3} and~\ref{propcaus4} complete the proof of Theorem~\ref{thmcausal}, except
for the cases where the curve $\gamma$ is null (or partially null).
Those particular cases can be treated by the same kind of arguments as above using transitivity of the
causal order and continuity.

Let us stress the importance of using the maximal cone of causal elements.
Indeed, following the results in~\cite{F6}, one can def\/ine a~partial order structure on the space of
states using only a~subcone of $\mathcal{C}$, as long as the condition $\overline{\Span_{{\mathbb
C}}(\mathcal{C})} = \overline{\widetilde{\mathcal{A}}}$ is still respected for some suitable unitisation
$\widetilde{\mathcal{A}}$.
For example, in the model considered in this section, if we use the subcone def\/ined by the causal
elements described in Lemma~\ref{lemmaC}, the global causal structure would be similar, but with
a~dif\/ferent constraint on $l(\gamma)$, allowing for more causal relations.
Indeed, one can check that the causal function $\mathbf{a}$ used in the proof of
Proposition~\ref{propcaus4} does not belong to such a~subcone.
Since in the commutative case the usual causal structure is recovered by using the maximal cone of causal
functions~\cite{F6}, the maximal cone should also be favoured in the noncommutative case.

\subsection{Causal structure for a~class of mixed states}
\label{secms}

As mentioned in Section~\ref{General}, the space of general states on $\mathcal{A}$ may contain
states that mix the space-time with the internal degrees of freedom.
The physical interpretation of such states is unclear and we shall not discuss it here.
On the other hand, there exists a~subclass of $S(\mathcal{A})$ of particular interest:
$P(\mathcal{A}_\mathcal{M}) \times S(\mathcal{A}_F) = \mathcal{M}(\widetilde{\mathcal{A}}_\mathcal{M})
\times S(\mathcal{A}_F)$, which we shall denote by $\mathcal{N}(\widetilde{\mathcal{A}})$.
In~\cite{IochKM} it has been shown that the Riemannian distance in the space of states on $P(M_2({\mathbb
C})) \cong S^2$ is equal to the Euclidean distance between the corresponding points of $S^2$, whenever the
two states in consideration have the same latitude.
This means that the shortest path between two pure states leads necessarily through mixed states.
Surprisingly enough, this is no longer the case in our model.
Note that the path of states $[0,T] \ni t \mapsto \chi(t)$ used in the proof of Proposition~\ref{propcaus3}
leads entirely through the space of pure states.
The purpose of this section is to show that indeed there is no ``short-cut'' through the space of
internal mixed states.

Let us recall that a~mixed state can always be written as a~convex combination of two pure states.
Since $P(\mathcal{A}_F) \cong S^2$, we have $S(\mathcal{A}_F) \cong B^2$, i.e.~a solid 2-dimensional ball
of radius one.
Again, let us stress that the isomorphism $S(\mathcal{A}_F) \cong B^2$ is only set-theoretic, the metric
aspect of~$B^2$ is ruled by the noncommutative geometry of $M_2({\mathbb C})$.
It is convenient to express the mixed states in terms of density matrices rather than convex combination.
Any mixed state in the subclass $\mathcal{N}(\widetilde{\mathcal{A}})$ can be written as $\omega_{p,\rho}$
with $\omega_{p,\rho}(a) = \Tr \rho  a(p) $.
The density matrices $\rho$ need to satisfy $\Tr \rho = 1$ and are conveniently parametrised~\cite{Zyczkowski} by
\begin{gather*}
\rho=\tfrac{1}{2}\left(1+\vec{r}\cdot\vec{\sigma}\right),
\qquad
\text{with}
\quad
\vec{r}\in B^2\subset{\mathbb R}^3
\quad
\text{and}
\quad
\vec{\sigma}=(\sigma_x,\sigma_y,\sigma_z),
\end{gather*}
where $\sigma_i$ are Pauli matrices.

We shall f\/irst write down an analogue of the Proposition~\ref{propcondcausal}.
\begin{proposition}
%\label{propcondcausalM}
Let us take two mixed states $\omega_{p,\rho},   \omega_{q,\sigma} \in \mathcal{N}(\widetilde{\mathcal{A}})$,
where $\rho$ and $\sigma$ are determined by the vectors $\vec{r},   \vec{s} \in B^2$
respectively.
Then $\omega_{p,\rho} \preceq \omega_{q,\sigma}$ if and only if\/~$\forall\, \mathbf{a} = \left(
\begin{smallmatrix}
a & -c
\\
-c^* & b
\end{smallmatrix}
\right) \in \mathcal{C}$,
\begin{gather*}
(1+s_z)a(q)-(1+r_z)a(p)+(1-s_z)b(q)-(1-r_z)b(p)
\\
\qquad
\geq2\Re\big\{s_x c(q)-r_x c(p)\big\}-2\Im\big\{s_y c(q)-r_y c(p)\big\}.
\end{gather*}
\end{proposition}
\begin{proof}
The Def\/inition~\ref{defcausal} implies that $\omega_{p,\rho} \preceq \omega_{q,\sigma}$ if and only if
$\forall\, \mathbf{a}\in\mathcal{C}$,
\begin{gather*}
\omega_{p,\rho}(\mathbf{a})\leq\omega_{q,\sigma}(\mathbf{a})
\quad
\Longleftrightarrow
\quad
\Tr\left(\rho \mathbf{a}(p)\right)\leq\Tr\left(\sigma \mathbf{a}(q)\right)
\\
\quad
\Longleftrightarrow
\quad
\tfrac{1}{2}\Tr
\begin{pmatrix}
1+r_z &  r_x-i r_y
\\
r_x+i r_y &  1-r_z
\end{pmatrix}
\begin{pmatrix}
a(p) &  -c(p)
\\
-c(p)^* &  b(p)
\end{pmatrix}
\\
\phantom{\Longleftrightarrow}
\quad-\tfrac{1}{2}\Tr
\begin{pmatrix}
1+s_z &  s_x-i s_y
\\
s_x+i s_y &  1-s_z
\end{pmatrix}
\begin{pmatrix}
a(q) &  -c(q)
\\
-c(q)^* &  b(q)
\end{pmatrix}
\leq0.
\tag*{\qed}
\end{gather*}
\renewcommand{\qed}{}
\end{proof}

The Propositions~\ref{Dfdeg},~\ref{propcaus1} and~\ref{propcaus2} are easily generalised using the same
elements of $\mathcal{C}$ as in the proofs for pure states:
\begin{proposition}
%\label{DfdegM}
If the finite-part Dirac operator is degenerate $($i.e.~$d_1 = d_2)$, then two mixed states
$\omega_{p,\rho},\omega_{q,\sigma} \in \mathcal{N}(\widetilde{\mathcal{A}})$ are causally related with
$\omega_{p,\rho} \preceq \omega_{q,\sigma}$ only if $p \preceq q$ in ${\mathbb R}^{1,1}$ and $\rho =
\sigma$.
\end{proposition}
\begin{proposition}
\label{propcaus1M}
Two states $\omega_{p,\rho},   \omega_{q,\sigma} \in \mathcal{N}(\widetilde{\mathcal{A}})$, with $\rho$ and
$\sigma$ associated with the vectors \mbox{$\vec{r},   \vec{s} \in B^2$} respectively, are causally related with
$\omega_{p,\rho} \preceq \omega_{q,\sigma}$ only if $p \preceq q$ in ${\mathbb R}^{1,1}$ and $r_z = s_z$.
\end{proposition}

The above Proposition encourages us to restrict the further investigation to the mixed states which have
the same latitude and change the parametrisation of the vectors $\vec{r}$ and $\vec{s}$ as follows:
\begin{gather*}
r_z=s_z=z,
\qquad
r_x\pm i r_y=r e^{\pm i\theta_r},
\qquad
s_x\pm i s_y=s e^{\pm i\theta_s}
\qquad
\text{with}
\quad
r, s\in\left[0,\sqrt{1-z^2}\right].
\end{gather*}
In this case, the condition for $\omega_{p,\rho} \preceq \omega_{q,\sigma}$ simplif\/ies to
\begin{gather}
\label{CausCondM}
(1+z)\left(a(q)-a(p)\right)+(1-z)\left(b(q)-b(p)\right)\geq2\Re\big\{c(q) s e^{i\theta_s}-c(p) r e^{i\theta_r}\big\}.
\end{gather}

The main theorem of this section is the following generalisation of the Theorem~\ref{thmcausal}.
\begin{theorem}
\label{thmcausal2}
Two states $\omega_{p,\rho},   \omega_{q,\sigma} \in \mathcal{N}(\widetilde{\mathcal{A}})$, with $\rho$ and
$\sigma$ associated with the vectors $\vec{r},   \vec{s} \in B^2$ respectively, are causally related with
$\omega_{p,\rho} \preceq \omega_{q,\sigma}$ if and only if the following conditions are respected:
\begin{itemize}\itemsep=0pt\sloppy
\item $p \preceq q$ in ${\mathbb R}^{1,1}$;
\item $r_z = s_z =z$;
\item $  l(\gamma)  \geq \sup_{\theta \in {\mathbb R}} \frac{\abs{\arccos\prt{\frac{s}{\sqrt{1-z^2}}
\cos(\theta_s+\theta)}-\arccos\prt{\frac{r}{\sqrt{1-z^2}} \cos(\theta_r+\theta)}}}{\abs{d_1-d_2}}$,
where $l(\gamma)$ represents the length of a~causal curve $\gamma$ going from $p$ to $q$ on ${\mathbb
R}^{1,1}$.
\end{itemize}
\end{theorem}
\begin{proof}
To prove the necessary condition for the causal relation to hold we make use of
Proposition~\ref{propcaus1M}.
Hence, it is suf\/f\/icient to show that $p \preceq q$, $r_z = s_z$ and $\omega_{p,\rho} \preceq
\omega_{q,\sigma}$ imply the existence of a~causal curve linking $p$ and $q$ with
\begin{gather*}
l(\gamma)\geq\frac{\big|\widetilde\theta_s-\widetilde\theta_r\big|}{d},
\qquad
\text{where}
\qquad
\widetilde\theta_{s|r}=\arccos\prt{\frac{s|r}{\sqrt{1-z^2}}\cos(\theta_{s|r}+\theta)},
\qquad
d=\abs{d_1-d_2}
\end{gather*}
and $\theta \in [0,2\pi[$ is chosen such that $\big|\widetilde\theta_s-\widetilde\theta_r\big|$ is maximal.
The argument by contradiction used in the proof of Proposition~\ref{propcaus4} can be applied, so we repeat
the same reasoning and stress only the dif\/ferences.
Since the argument of arccos in the def\/inition of $\widetilde\theta_{s|r}$ is in the interval $[-1,1]$,
we have $\widetilde\theta_s,\widetilde\theta_r \in [0,\pi]$.
We shall consider $\widetilde\theta_s,\widetilde\theta_r \in ]0,\pi[$ and the limiting cases can be dealt
with by the argument of transitivity of the causal order as in Proposition~\ref{propcaus4}.

We assume that $l(\gamma) < \frac{\abs{ \widetilde\theta_s-\widetilde\theta_r}}{d}$ for every future
directed timelike curve $\gamma$ with $\gamma(0)=p$ and $\gamma(T)=q$ and we show that there exists
a~causal element $\mathbf{a} = \left(
\begin{smallmatrix}
a & -c
\\
-c^* & b
\end{smallmatrix}
\right) \in \mathcal{C}$ such that:
\begin{gather*}
%\label{antigoal2}
(1+z)\left(a(q)-a(p)\right)+(1-z)\left(b(q)-b(p)\right)<2\Re\big\{c(q) s e^{i\theta_s}-c(p) r e^{i\theta_r}\big\}.
\end{gather*}
The of\/f-diagonal terms are def\/ined as:
\begin{gather*}
c(\gamma(t))=-\csc(d l_\gamma(t)+\epsilon)e^{i\theta_c},
\qquad
\text{where}
\quad
\begin{cases}
\epsilon=\widetilde\theta_r,
&
\theta_c=\theta
\quad
\text{if}
\quad
\widetilde\theta_s-\widetilde\theta_r>0,
\\
\epsilon=-\widetilde\theta_r+\pi,
&
\theta_c=\theta+\pi
\quad
\text{if}
\quad
\widetilde\theta_s-\widetilde\theta_r<0.
\end{cases}
\end{gather*}
The diagonal terms are def\/ined as in the formulas~\eqref{diagel} using $\abs{\varphi_1}=\sqrt{1+z}$ and
$\abs{\varphi_2}=\sqrt{1-z}$.
Then the equality corresponding to~\eqref{expr1} is
\begin{gather}
(1+z)\left(a(q)-a(p)\right)+(1-z)\left(b(q)-b(p)\right)=2\sqrt{1-z^2}
\prt{-\cot(d l_\gamma(T)+\epsilon)+\cot(\epsilon)}
\label{expr3}
\end{gather}
and since $(s|r) \cos(\theta_{s|r}+\theta_c) = \sgn(\widetilde\theta_s-\widetilde\theta_r) \sqrt{1-z^2}
\cos\big(\pm \widetilde\theta_{s|r}\big)$ we have
\begin{gather}
2\Re\big\{c(q) s e^{i\theta_s}-c(p) r e^{i\theta_r}\big\}
=2\prt{-\frac{s\cos(\theta_s+\theta_c)}{\sin(d l_\gamma(T)+\epsilon)}
+\frac{r\cos(\theta_r+\theta_c)}{\sin(\epsilon)}}
\nonumber
\\
\hphantom{2\Re\big\{c(q) s e^{i\theta_s}-c(p) r e^{i\theta_r}\big\}}{}
 =2\sqrt{1-z^2}\prt{-\frac{\sgn(\widetilde\theta_s-\widetilde\theta_r)\cos(\pm\widetilde\theta_s)}
{\sin(d l_\gamma(T)+\epsilon)}+\cot(\epsilon)}.
\label{expr4}
\end{gather}
By comparing~\eqref{expr3} and~\eqref{expr4} it remains to prove that
\begin{gather*}
\cos(d l_\gamma(T)+\epsilon)>\sgn(\widetilde\theta_s-\widetilde\theta_r)\cos(\pm\widetilde\theta_s).
\end{gather*}
Since cosine is decreasing on $]0,\pi[$ we have:
\begin{itemize}\itemsep=0pt
 \item If $\widetilde\theta_s-\widetilde\theta_r > 0$,
we get $\cos( d l_\gamma(T)+\widetilde\theta_r) > \cos(\widetilde\theta_s)$
which is true since $0 < d l_\gamma(T)+\widetilde\theta_r <
\big|\widetilde\theta_s-\widetilde\theta_r\big|+\widetilde\theta_r =\widetilde\theta_s < \pi$.
\item If $\widetilde\theta_s-\widetilde\theta_r < 0$, we get $\cos( d l_\gamma(T)-\widetilde\theta_r
+\pi) > \cos(-\widetilde\theta_s + \pi)$ which is true since $0 < d l_\gamma(T)-\widetilde\theta_r + \pi
< \big|\widetilde\theta_s-\widetilde\theta_r\big|-\widetilde\theta_r + \pi =-\widetilde\theta_s + \pi <\pi$.
\end{itemize}

So the necessary condition is completed.
In order to prove the suf\/f\/icient condition, we use the proof of Proposition~\ref{propcaus3} with the
following changes:
\begin{itemize}
\itemsep=0pt \item $l(\gamma) = \frac{\abs{ \widetilde\theta_s-\widetilde\theta_r}}{d}$; \item
$\xi_1=\varphi_1=\sqrt{1+z}$; \item $\chi(t) = \sqrt{1-z}  e^{i(\widetilde\theta_r +
\sgn(\widetilde\theta_s-\widetilde\theta_r)  d  l_\gamma(t))}$, which implies $\abs{\chi(t)}= \sqrt{1-z}
$ $\forall\, t\in[0,T]$, $\chi(0)=\sqrt{1-z}  e^{i\widetilde\theta_r} $ and $\chi(T)=\sqrt{1-z}
 e^{i\widetilde\theta_s}$.
\item For an arbitrary $\mathbf{a} = \left(\begin{smallmatrix}a & -c\\-c^* & b\end{smallmatrix}
\right) \in \mathcal{C}$ we consider instead an associate causal element
$\widetilde{\mathbf{a}} = \left(\begin{smallmatrix}a & -\widetilde c\\-\widetilde c^* & b\end{smallmatrix}
\right) \in \mathcal{C}$ with $\widetilde c = c  e^{-i\theta}$.
One can check using Proposition~\ref{propC}$(c)$ with $\widetilde \alpha_3 = \alpha_3  e^{-i\theta}$ and
$\widetilde \alpha_4 = \alpha_4  e^{-i\theta}$ that $ \mathbf{a} \in \mathcal{C}
\
\Leftrightarrow
\
\widetilde{\mathbf{a}} \in \mathcal{C}$.
\end{itemize}

By repeating the proof of Proposition~\ref{propcaus3} until~\eqref{goto}, we get:
\begin{gather*}
(1+z)\left(a(q)-a(p)\right)+(1-z)\left(b(q)-b(p)\right)\geq2\sqrt{1+z}
 \Re\big\{\chi(T) \widetilde c(q)-\chi(0) \widetilde c(p)\big\}.
\end{gather*}

From the formula~\eqref{CausCondM}, it remains to show that the following inequality is true for at least
one $\theta \in [0,2\pi)$:
\begin{gather}
\label{goalmixed}
2\sqrt{1+z} \Re\big\{\chi(T) \widetilde c(q)-\chi(0) \widetilde c(p)\big\}\geq2\Re\big\{c(q) s e^{i\theta_s}-c(p) r e^{i\theta_r}\big\}.
\end{gather}

Let us set $c(\gamma(t)) = \abs{c(\gamma(t))} e^{i\theta_c(t)}$.
At f\/irst let us assume that $\theta_c(0)=\theta_c(T)$.
Then we can choose $\theta=\theta_c(0)=\theta_c(T)$ which implies $\widetilde c(p) = \abs{c(p)} \in
{\mathbb R}$ and $\widetilde c(q) = \abs{c(q)} \in {\mathbb R}$.
Hence,
\begin{gather*}
2\sqrt{1+z} \Re\big\{\chi(T) \widetilde c(q)-\chi(0) \widetilde c(p)\big\}
=2\abs{c(q)}\sqrt{1-z^2} \cos(\widetilde\theta_s)-2\abs{c(p)}\sqrt{1-z^2} \cos(\widetilde\theta_r)
\\
\qquad
=2\abs{c(q)}s \cos(\theta_s+\theta)-2\abs{c(p)}r \cos(\theta_r+\theta)
\\
\qquad
=2\Re\big\{\abs{c(q)}e^{i\theta} s e^{i\theta_s}-\abs{c(p)}e^{i\theta} r e^{i\theta_r}\big\}
=2\Re\big\{c(q) s e^{i\theta_s}-c(p) r e^{i\theta_r}\big\}.
\end{gather*}

Let us now assume that $\theta_c(0) \neq \theta_c(T)$. The l.h.s.\ of~\eqref{goalmixed} can be expanded as:
\begin{gather}
2\sqrt{1+z} \Re\big\{\chi(T) \widetilde c(q)-\chi(0) \widetilde c(p)\big\}
\nonumber
\\
\qquad
=2\sqrt{1-z^2}\abs{c(q)}\cos(\theta_c(T)-\theta+\widetilde\theta_s)-2\sqrt{1-z^2}\abs{c(p)}
\cos(\theta_c(0)-\theta+\widetilde\theta_r)
\nonumber
\\
\qquad
=2\abs{c(q)}\cos(\theta_c(T)-\theta) s\cos(\theta_s+\theta)-2\sqrt{1-z^2}\abs{c(q)}
\sin(\theta_c(T)-\theta)\sin(\widetilde\theta_s)
\nonumber
\\
\qquad\phantom{=}
{}-2\abs{c(p)}\cos(\theta_c(0)-\theta) r\cos(\theta_r+\theta)+2\sqrt{1-z^2}\abs{c(p)}
\sin(\theta_c(0)-\theta)\sin(\widetilde\theta_r)
\label{dvlp1}
\end{gather}
and the r.h.s.\ as:
\begin{gather}
2\Re\big\{c(q) s e^{i\theta_s}-c(p) r e^{i\theta_r}\big\}
=2\Re\big\{c(q)e^{-i\theta} s e^{i(\theta_s+\theta)}-c(p)e^{-i\theta} r e^{i(\theta_r+\theta)}\big\}
\nonumber
\\
\qquad
=2\abs{c(q)}\cos(\theta_c(T)-\theta) s\cos(\theta_s+\theta)-2\abs{c(q)}
\sin(\theta_c(T)-\theta)s\sin(\theta_s+\theta)
\nonumber
\\
\qquad\phantom{=}
{}-2\abs{c(p)}\cos(\theta_c(0)-\theta) r\cos(\theta_r+\theta)+2\abs{c(p)}
\sin(\theta_c(0)-\theta)r\sin(\theta_r+\theta).
\label{dvlp2}
\end{gather}

If we compare the two expressions~\eqref{dvlp1} and~\eqref{dvlp2}, in order to respect the
inequality~\eqref{goalmixed}, we need to prove that there exists at least one $\theta\in{\mathbb R}$ such
that:
\begin{gather}
-2\sqrt{1-z^2}\abs{c(q)}\sin(\theta_c(T)-\theta)\sin(\widetilde\theta_s)+2\sqrt{1-z^2}\abs{c(p)}
\sin(\theta_c(0)-\theta)\sin(\widetilde\theta_r)
\nonumber
\\
\phantom{\Longleftrightarrow\quad}
\geq-2\abs{c(q)}\sin(\theta_c(T)-\theta)s\sin(\theta_s+\theta)+2\abs{c(p)}
\sin(\theta_c(0)-\theta)r\sin(\theta_r+\theta)
\nonumber
\\
\Longleftrightarrow\quad
\abs{c(p)}\sin(\theta_c(0)-\theta)\left[\sqrt{1-z^2}\sin(\widetilde\theta_r)-r\sin(\theta_r+\theta)\right]
\nonumber
\\
\phantom{\Longleftrightarrow\quad}
\geq\abs{c(q)}\sin(\theta_c(T)-\theta)\left[\sqrt{1-z^2}\sin(\widetilde\theta_s)-s\sin(\theta_s+\theta)\right].
\label{rest}
\end{gather}

By using the formula $\sin(\arccos(x))=\sqrt{1-x^2}$ and $s,r \leq \sqrt{1-z^2}$, we have that:
\begin{gather*}
\begin{split}
& \sqrt{1-z^2}\sin(\widetilde\theta_{s|r})=\sqrt{(1-z^2)-(s|r)^2\cos(\theta_{s|r}+\theta)}
\\
& \qquad
\geq\sqrt{(s|r)^2-(s|r)^2\cos(\theta_{s|r}+\theta)}=(s|r)\abs{\sin(\theta_{s|r}+\theta)}.
\end{split}
\end{gather*}
So the two brackets inside~\eqref{rest} are always non-negative, and the inequality is manifestly respected
when $\sin(\theta_c(0)-\theta) > 0$ and $\sin(\theta_c(T)-\theta) < 0$.
Since $\theta_c(0) \neq \theta_c(T)$, there always exists one $\theta\in{\mathbb R}$ making the f\/irst
expression positive and the second negative.
Therefore, the proof is complete.
\end{proof}

The constraint for mixed states can be given a~geometrical interpretation.
The dif\/ference between the arccosine values in $\abs{\widetilde\theta_s-\widetilde\theta_r}$ is the
length of a~projection of the segment $\vec s-\vec r$ on the parallel of radius $\sqrt{1-z^2}$.
The supremum over the angle $\theta$ means that the projection should always be the maximal one.
This constraint is thus compatible with the formula~\eqref{speedoflight} for the pure states derived in
Section~\ref{secps}.
It also leads to the conclusion that paths going trough the space of mixed states
$\mathcal{N}(\widetilde{\mathcal{A}})$ -- in particular the straight line path corresponding to the
realisation of the Riemannian distance in the internal space~\cite{IochKM} -- do not shorten the amount of
proper time needed for the path to be causal.
Hence, a~unitary evolution (i.e.~the one involving only pure states) in the internal space is, in a~sense,
favoured by the system.

\section{Conclusions}

In this paper, we have provided a~def\/inition of an \emph{almost commutative Lorentzian spectral triple}
which describes a~product of a~globally hyperbolic Lorentzian manifold with a~f\/inite noncommutative space.
Following the results in~\cite{F6}, we equipped the space of states of this spectral triple with a~partial
order relation that generalises the causal structure of Lorentzian manifolds.
After some general results we explored in details the causal structure of a~simple almost commutative
space-time based on the algebra $\mathcal{S}({\mathbb R}^{1,1}) \otimes M_2({\mathbb C})$.
We have shown that the space of pure states is isomorphic to the set ${\mathbb R}^{1,1} \times S^2$, with
the geometry of the $S^2$ component governed by a~noncommutative matrix algebra.
Our main result~-- Theorem~\ref{thmcausal}~-- shows that the causal structure of this simple almost
commutative space-time is highly non-trivial.
Its main features are the following:
\begin{enumerate}\itemsep=0pt
\item No violation of the classical causality relations in the 2-dimensional Minkowski
space-time component occurs.
\item The Riemannian noncommutative geometry of $M_2({\mathbb C})$ represented on ${\mathbb C}^2$ is
respected, i.e.~the parallels of $S^2$ are causally disconnected.
\item The admissible motions in the Minkowski space-time and the internal space $S^2$ are intrinsically
related by an explicit formula~\eqref{speedoflight}.
\item The path realising the shortest Riemannian distance between two pure states, which leads through the
space of mixed states, is equivalent, from the causal point of view, to a~path enclosed entirely in the
space of pure states.
\end{enumerate}

Naively, one could think that in the setting of almost commutative geometries the motion in the internal
space should be completely independent of that in the base space-time.
On the contrary, the simple model we have chosen to investigate has revealed a~lot of interesting and
unexpected features of the causal relations in the space of states.
This shows that the causal structure of general almost commutative Lorentzian spectral triples dif\/fers
signif\/icantly from the one of Lorentzian manifolds and its exploration may lead to surprising conclusions.

The illustrative example presented in Section~\ref{secps} can be generalised to curved manifolds.
One can use Proposition~\ref{conformal} to extend the results of Theorem~\ref{thmcausal} to a~general
globally hyperbolic Lorentzian metric on ${\mathbb R}^{2}$, since such a~metric is always locally
conformally f\/lat~\cite{Nakahara}.
On the other hand, this feature is no longer true when the dimension of space-time is larger then 2.
Exploring space-times of larger dimensions is also technically more dif\/f\/icult and does not necessarily
lead to a~linear characterisation of the causal elements as given in Proposition~\ref{propC}.

In addition to the possible generalisations of the space-time component it would be desirable to
investigate the causal structure of almost commutative geometries with other matrix algebras and direct
sums of the latter.
Here the main problem, apart from the increasing level of complexity of the formulas for higher dimensional
matrices, is that the Riemannian noncommutative geometry of the associated space of states has not been
studied in details.

Finally, let us comment on possible physical applications of the results contained in this paper.
As stressed in the Introduction, almost commutative geometries provide a~f\/irm mathematical framework for
the study of fundamental interactions.
Although the physical interpretation of the space of states of such geometries (especially the mixed
states) is not clear, our investigations show that one can consider the motion in internal space in
a~meaningful way.
What is more, the possibility of the internal movement turns out to be dependent on the speed assumed in
the space-time component.
It suggests that the ``speed of light constraint'' in this model should be understood in terms of resultant
space-time and internal velocities, with a~proportionality parameter determined by the f\/inite-part Dirac
operator $D_F$.
Certainly, it would be interesting to apply the presented results in some concrete physical model, with the
standard model of fundamental interactions as an ultimate goal.

\appendix \section*{Appendix}

In this Appendix, we show that the function $\mathbf{a}$ def\/ined in the proof of
Proposition~\ref{propcaus4} (and also used in the proof of Theorem~\ref{thmcausal2}) is a~causal function
along the curve $\gamma$, i.e.~that the matrix given in Proposition~\ref{propC}$(b)$ is positive
semi-def\/inite.

Let us compute the entries of that matrix.
We have:
\begin{itemize}\itemsep=0pt
\item $a_{,0} + a_{,1} = d \sqrt{\frac{\lambda_2}{\lambda_1}}
\frac{\abs{\varphi_2}}{\abs{\varphi_1}} \csc^2\Theta$, \item $a_{,0}-a_{,1} =
d \sqrt{\frac{\lambda_1}{\lambda_2}} \frac{\abs{\varphi_2}}{\abs{\varphi_1}} \csc^2\Theta$, \item $b_{,0}
+ b_{,1} = d \sqrt{\frac{\lambda_2}{\lambda_1}} \frac{\abs{\varphi_1}}{\abs{\varphi_2}} \csc^2\Theta$,
\item $b_{,0}-b_{,1} = d \sqrt{\frac{\lambda_1}{\lambda_2}} \frac{\abs{\varphi_1}}{\abs{\varphi_2}}
\csc^2\Theta$, \item $c = -\csc\Theta e^{i \theta_c}$,
\end{itemize}
with $\Theta(t)=d l_\gamma(t) + \epsilon$ and $\theta_c=\sgn(\theta_\varphi-\theta_\xi)\epsilon
-\theta_\xi$ and where the parameters $t$ and $\gamma(t)$ have been omitted.

The partial derivatives of $c$ are def\/ined in such a~way that its directional derivative is maximal along
$\gamma$ and corresponds to
\begin{gather*}
\dv{}{t}(c\circ\gamma)=-d \dv{l_\gamma}{t}\cos\Theta\csc^2\Theta e^{i\theta_c}
=-d 2\dot\gamma_0\sqrt{\lambda_1\lambda_2}\cos\Theta\csc^2\Theta e^{i\theta_c}.
\end{gather*}
This can be done by setting:{\samepage
\begin{itemize}
\itemsep=0pt \item $c_{,0} + c_{,1} =-d \sqrt{\frac{\lambda_2}{\lambda_1}} \cos\Theta\csc^2\Theta  e^{i
\theta_c}$, \item $c_{,0}-c_{,1} =-d \sqrt{\frac{\lambda_1}{\lambda_2}} \cos\Theta\csc^2\Theta  e^{i
\theta_c}$,
\end{itemize}
since $\dv{}{t}(c\circ\gamma) = \dot\gamma_0 \prt{\lambda_1 (c_{,0} + c_{,1}) + \lambda_2 (c_{,0}-c_{,1})
}.$}

So we must prove that the following matrix is positive semi-def\/inite:{\samepage
\begin{gather*}
A=d \csc^2\Theta\left(
\begin{matrix}
\sqrt{\frac{\lambda_2}{\lambda_1}}\frac{\abs{\varphi_2}}{\abs{\varphi_1}}
 &  0 &  \sqrt{\frac{\lambda_2}{\lambda_1}}\cos\Theta e^{i\theta_c} &  \pm\sin\Theta e^{i\theta_c}
\\
0 &  \sqrt{\frac{\lambda_1}{\lambda_2}}\frac{\abs{\varphi_2}}{\abs{\varphi_1}}
 &  \mp\sin\Theta e^{i\theta_c} &  \sqrt{\frac{\lambda_1}{\lambda_2}}\cos\Theta e^{i\theta_c}
\\
\sqrt{\frac{\lambda_2}{\lambda_1}}\cos\Theta e^{-i\theta_c} &  \mp\sin\Theta e^{-i\theta_c}
 &  \sqrt{\frac{\lambda_2}{\lambda_1}}\frac{\abs{\varphi_1}}{\abs{\varphi_2}} &  0
\\
\pm\sin\Theta e^{-i\theta_c} &  \sqrt{\frac{\lambda_1}{\lambda_2}}\cos\Theta e^{-i\theta_c}
 &  0 &  \sqrt{\frac{\lambda_1}{\lambda_2}}\frac{\abs{\varphi_1}}{\abs{\varphi_2}}
\end{matrix}
\right),
\end{gather*}
where $\pm$ corresponds to $\sgn(d_1-d_2)$.}

Since the eigenvalues of $A$ are the roots of the characteristic polynomial:
\begin{gather*}
\det(A-\lambda1)=\lambda^4-c_1\lambda^3+c_2\lambda^2-c_3\lambda+c_4,
\end{gather*}
from Vieta's formulas it is suf\/f\/icient to check that $c_k \geq 0$ for $k=1,\dots,4$.
We can notice that this matrix is singular.
Indeed, by multiplying the f\/irst line by $\frac{\abs{\varphi_1}}{\abs{\varphi_2}} \cos\Theta  e^{-i
\theta_c}$ and the second line by $\mp\sqrt{\frac{\lambda_2}{\lambda_1}}
\frac{\abs{\varphi_1}}{\abs{\varphi_2}} \sin\Theta  e^{-i \theta_c}$ and adding them, we get exactly the
third line.
So $c_4 = \det(A) = 0$.

The other coef\/f\/icients can be calculated by algebraic computations from Newton's identities using the
following traces:
\begin{gather*}
\tr A=d \csc^2\Theta\frac{1}{\sqrt{\lambda_1}\sqrt{\lambda_2}\abs{\varphi_1}\abs{\varphi_2}},
\\
\tr A^2=d^2\csc^4\Theta\frac{-2\abs{\varphi_1}^2\abs{\varphi_2}
^2(\lambda_2-\lambda_1)^2\sin^2\Theta-2\lambda_1\lambda_2+1}{\prt{\sqrt{\lambda_1}\sqrt{\lambda_2}
\abs{\varphi_1}\abs{\varphi_2}}^2},
\\
\tr A^3=d^3\csc^6\Theta\frac{-3\abs{\varphi_1}^2\abs{\varphi_2}
^2(\lambda_2-\lambda_1)^2\sin^2\Theta-3\lambda_1\lambda_2+1}{\prt{\sqrt{\lambda_1}\sqrt{\lambda_2}
\abs{\varphi_1}\abs{\varphi_2}}^3},
\\
c_1=\tr A=d \csc^2\Theta\frac{1}{\sqrt{\lambda_1}\sqrt{\lambda_2}\abs{\varphi_1}\abs{\varphi_2}}\geq0,
\\
c_2=\frac12\prt{\prt{\tr A}^2-\tr A^2}=d^2\csc^4\Theta\frac{\abs{\varphi_1}^2\abs{\varphi_2}
^2(\lambda_2-\lambda_1)^2\sin^2\Theta+\lambda_1\lambda_2}{\prt{\sqrt{\lambda_1}\sqrt{\lambda_2}
\abs{\varphi_1}\abs{\varphi_2}}^2}\geq0,
\\
c_3=\frac16\prt{\prt{\tr A}^3-3\tr A^2\tr A+2\tr A^3}=0.
\end{gather*}

So all of the coef\/f\/icients of the characteristic polynomial are non-negative and the matrix is positive
semi-def\/inite.

\subsection*{Acknowledgements}

This work was supported by a~grant from the John Templeton Foundation.

\pdfbookmark[1]{References}{ref}
\LastPageEnding

\end{document}